\documentclass[12pt]{article}
\usepackage{amsmath,amsfonts,latexsym,amssymb,amsthm,graphicx,subcaption}

\usepackage{booktabs} 
\usepackage{multirow}
\usepackage{makecell}
\usepackage{placeins}
\usepackage{caption}
\usepackage{float}

\usepackage{bm}
\usepackage[round]{natbib}
\usepackage{hyperref}
\usepackage[margin=1in]{geometry}
\usepackage{setspace}
\usepackage[normalem]{ulem}
\usepackage[round]{natbib}
\usepackage{authblk}

\newtheorem{proposition}{Proposition}
\theoremstyle{definition}

\def\SCS{\widehat{\mathcal{S}}_{\alpha}}
\def\LB{\underline{\SCS}}
\def\Scal{{\mathcal{S}}}
\def\hats0{{\hat {\mathbf{s}}_0}}
\def\bfs{{\mathbf{s}}}

\def\bfV{{\mathbf{V}}}
\def\bf0{{\mathbf{0}}}
\def\bftheta{{\boldsymbol{\theta}}}

\title{Selection Confidence Sets for Equally Weighted Portfolios}

\author[2]{Davide Ferrari}
\author[1]{Alessandro Fulci\thanks{Corresponding author: E-mail: \texttt{alessandro.fulci@unitn.it}. Address: Department of Economics and Management, University of Trento, Via Inama 5, 38122 Trento, Italy}}
\author[1]{Sandra Paterlini}

\affil[1]{Department of Economics and Management, University of Trento}
\affil[2]{Faculty of Economics and Management, Free University of Bozen-Bolzano}

\date{\today}

\begin{document}

\doublespacing

\maketitle

\begin{abstract}
Given a universe of $N$ assets, investors often form equally weighted portfolios (EWPs) by selecting subsets of assets. EWPs are simple, robust, and competitive out-of-sample, yet the uncertainty about which subset truly performs best is largely ignored. Traditional approaches typically rely on a single selected portfolio, but this fails to consider alternative investment strategies that may perform just as well when accounting for statistical uncertainty.  To address this selection uncertainty, we introduce the Selection Confidence Set (SCS) for EWPs: the set of all portfolios that, under a given loss function and at a specified confidence level, contains the unknown set of optimal portfolios under repeated sampling. The SCS quantifies selection uncertainty by identifying a range of plausible portfolios, challenging the idea of a uniquely optimal choice. Like a confidence set, its size reflects uncertainty -- growing with noisy or limited data, and shrinking as the sample size increases.  Theoretically, we establish that the SCS covers the unknown optimal selection with high probability and characterize how its size grows with underlying uncertainty, corroborating these results through Monte Carlo experiments. Applications to the French 17-Industry Portfolios and Layer-1 cryptocurrencies underscore the importance of accounting for selection uncertainty when comparing equally weighted strategies.
\end{abstract}

Keywords: Equally Weighted Portfolios, Selection Confidence Set, Selection Uncertainty, Subset Selection, Wald Test.

\section{Introduction}\label{sec1}

The Mean-Variance theory developed by \cite{M.52} is the cornerstone of portfolio theory as it provides a rigorous mathematical framework for identifying portfolios that maximize expected return for a given level of risk. Despite subsequent advances in portfolio selection (e.g., \cite{S.64, R.76, FF.92, E.04, GO.15, C.21}), a central challenge remains the uncertainty surrounding the asset selection process. Even small estimation errors in expected returns or covariances can lead to large variability in the chosen allocation, making it difficult to deem any single portfolio superior. This sensitivity can lead to high turnover \citep{C.13} and make allocations untradeable once transaction costs are accounted for \citep{DM.09}. Numerous studies have sought to mitigate the effects of estimation error from various perspectives, including  Bayesian approaches, shrinkage estimators,  factor-based models, and the combination of tangent, minimum-variance, and risk-free portfolios (e.g., see \cite{DM.13}, \cite{B.21}, \cite{BO.22},  \cite{BO.24}, \cite{Bod.25}). 

A growing body of research shows that simple allocation rules can outperform optimized strategies. Particularly, the Equally Weighted Portfolio (EWP), or $1/N$ rule, stands out as a robust benchmark due to its simplicity, strong out-of-sample performance, resilience to estimation error, and built-in diversification \citep{DM.09, P.15, Bod.22}. By avoiding estimation of return moments,  EWPs sidestep
instability, limit shorting, and exploit mean-reversion when rebalancing \citep{M.17}. Motivated by these advantages, recent work has explored data-driven methods for selecting optimal EWP subsets. \cite{L.20} propose a deep learning algorithm that identifies top-performing assets, yielding substantial out-of-sample gains. \cite{CA.24} apply tree-based classifiers to select equally weighted subsets of S\&P 500 constituents, achieving superior risk-adjusted performance relative to the full EWP. Other learning approaches -- such as support vector machines, gradient boosting, and nature-inspired heuristics  -- have also been investigated, but their effectiveness is often limited by noise and the existence of many similarly performing portfolios, making it difficult to identify a single dominant EWP (e.g., see \cite{P.19}, \cite{C.21}, \cite{A.21}).

Motivated by the need to characterize uncertainty in EWP selection, this paper adopts a multi-portfolio perspective rather than focusing on identifying a single best portfolio.  To formalize selection uncertainty, we introduce the Selection Confidence Set (SCS) for EWPs: the set of all asset selections whose performance under a chosen loss function is statistically indistinguishable from that of the unknown optimal EWP at the  $(1-\alpha)\%$ confidence level, with $0<\alpha<1$. To construct the SCS, we use a Wald-type screening test to retain all the feasible portfolios not distinguishable from the unknown optimum. Under regularity conditions, the SCS achieves asymptotic coverage of at least $1-\alpha$, offering a principled basis for multi-portfolio selection and quantifying uncertainty in portfolio decisions.

\begin{figure}[ht]
    \centering
    \includegraphics[width=0.5\textwidth]{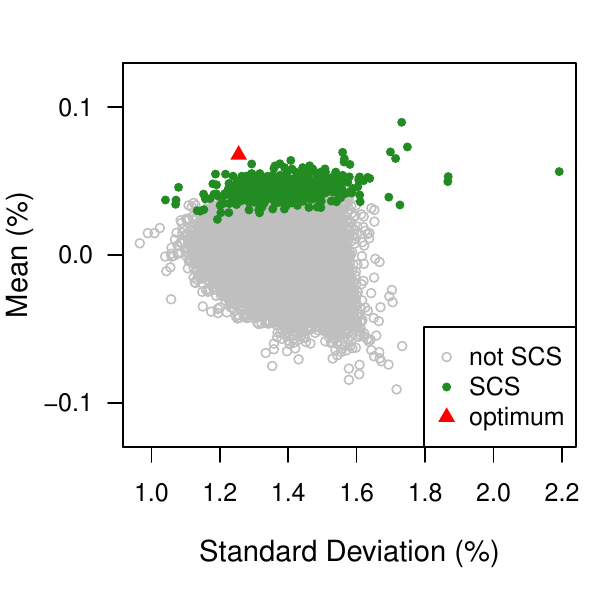}
    \caption{Sample means and standard deviations of EWPs from the French 17-Industry Portfolios data over a 3-year window. The red triangle marks the empirical optimum minimizing \(L(\mu,\sigma^2) = \sigma^2 - 0.5\mu\). Solid green dots represent portfolios in the 95\% SCS based on the Wald-type test in Section~\ref{sec:Methods}; open gray dots indicate portfolios outside the SCS.}
  \label{fig:Front}
\end{figure}

Analogous to confidence intervals in parameter estimation, the SCS offers a statistical framework for quantifying uncertainty in portfolio selection by identifying statistically plausible candidates. It allows practitioners to assess whether a proposed portfolio is empirically supported, helping to avoid overconfident or unwarranted choices. The size and composition of the SCS serve as diagnostics for selection uncertainty: large sets reflect limited information and difficulty in distinguishing among portfolios, often due to noise or small samples. Empirically, we find that an unexpectedly large number of EWPs are statistically indistinguishable from the empirical optimum in common situations, revealing substantial uncertainty often overlooked in practice. As sample size increases, the SCS naturally concentrates around the true optimum. As an illustration, 
Figure~\ref{fig:Front}  shows the sample mean and standard deviation of EWP returns over a 3-year window for the French 17-Industry Portfolios data described in Section~\ref{sec:examples}. At the 5\% significance level, 408 portfolios have mean-variance losses \(L(\mu,\sigma^2) = \sigma^2 - 0.5\mu\) statistically indistinguishable from the empirical optimum based on the test in Section~\ref{sec:Methods}; these portfolios form the SCS. In contrast, the remaining portfolios perform significantly worse. The size and diversity of the SCS underscore the extent of selection uncertainty and the risk of overinterpreting any single estimated optimum.

 While confidence sets for model selection have been studied in statistics, this paper is the first to develop a practical inferential framework for equally weighted financial portfolios. \cite{FY.15} introduced selection confidence sets for linear regression using F-test screening, and \cite{Z.19} extended the approach to general parametric models via likelihood ratio tests. Previously, \cite{H.11} proposed the Model Confidence Set (MCS), which retains models that are not significantly worse than the best in-sample performer while controlling for multiple comparisons. This method has been widely applied in empirical finance (e.g., \cite{AM.20}, \cite{LI.22}, \cite{CH.23}). While sharing the same multi-model rationale as the SCS, the MCS does not necessarily aim to cover the population-optimal model. In contrast, our SCS is inferential in nature and aligns with the approach of \cite{FY.15}, targeting coverage of the true optimal set of portfolios in the population with a specified confidence level under repeated sampling. 

The rest of the  paper is organized as follows. Section~\ref{sec:Methods} introduces the EWP selection framework, the SCS construction via Wald test screening, and examples under independent sampling. Section~\ref{sec:properties} presents theoretical coverage guarantees and asymptotic size analysis. Section~\ref{sec:metrics} introduces post-selection metrics to summarize selection uncertainty and the composition of the SCS. Section~\ref{sec:Monte_Carlo} reports Monte Carlo results on coverage and size. Section~\ref{sec:examples} applies the method to the French 17-Industry Portfolios and Layer-1 cryptocurrency datasets. Section~\ref{sec:Conclusion} concludes. Proofs are provided in the Appendix, while additional numerical results are collected in the Supplementary Materials.

\section{Confidence Sets for the optimal EWPs}
\label{sec:Methods}

\subsection{Setup and Objectives}
\label{sec:setting}

Let $\boldsymbol{X} = (X_1, \dots, X_N)^\top \in \mathbb{R}^N$,    denote a random vector of asset returns, $N < \infty$ is the finite number of assets. A portfolio is a linear combination of the components of $\boldsymbol{X}$. We consider portfolios constructed from $N$ assets, defined as
\begin{equation}\label{eq:portfolio}
Y_{\mathbf{s}} = \dfrac{\sum_{i=1}^N s_i X_i}{\sum_{j=1}^N s_j}, \quad \mathbf{s} = (s_1, \dots, s_N) \in \mathcal{S}:=\{0,1\}^N \setminus \{\mathbf{0}\},
\end{equation}
where $\mathbf{s}$ is a binary vector indicating asset selection, i.e., $s_i = 1$ if asset $i$ is selected, and $s_i = 0$ otherwise, and $\mathcal{S}$ represents the set of all feasible selections. 
The  mean and variance of $Y_{\bfs}$ are denoted by $
\mu_{\mathbf{s}} := \mathbb{E}\left[ Y_{\mathbf{s}} \right]$ and 
$\sigma^2_{\mathbf{s}} := \text{var}\left[ Y_{\mathbf{s}} \right]$, respectively. While we focus on the general case where $\Scal$ includes all asset subset combinations, in practice restrictions may be introduced to reflect investment policies, transaction costs, or operational requirements.

We define the loss function as a continuously differentiable mapping $L: \mathbb{R} \times \mathbb{R}_{+} \to \mathbb{R}$,   $(\mu, \sigma^2) \mapsto L(\mu, \sigma^2)$, assigning a real-valued performance criterion to each portfolio based on its mean return $\mu$ and variance $\sigma^2$.
The loss incurred by portfolio \( Y_{\mathbf{s}} \) depends on the asset selection \( \mathbf{s} \) only through its mean and variance, \( \mu_{\mathbf{s}} \) and \( \sigma^2_{\mathbf{s}} \); for instance, under a mean–variance criterion, one may take  $
L(\mu_{\mathbf{s}}, \sigma^2_{\mathbf{s}}) = \sigma^2_{\mathbf{s}} - \gamma \mu_{\mathbf{s}}$ for some $ \gamma > 0$. Based on $L$, we define the set of optimal asset selections as  
\begin{equation}\label{eq:optimal}
\mathcal{S}_0 := \underset{\mathbf{s} \in \mathcal{S}}{\arg\min} \ L(\mu_{\mathbf{s}}, \sigma^2_{\mathbf{s}}).
\end{equation}
Since \( \Scal \) is finite, the minimum is always attained and \( \mathcal{S}_0 \) contains at least one selection. In the rest of the paper, we use \( L_0 := L(\mu_{\bfs_0}, \sigma^2_{\bfs_0}) \), \( \bfs_0 \in \Scal_0 \) to denote the minimum loss.

Given a sample \( \boldsymbol{X}^{(1)}, \dots, \boldsymbol{X}^{(T)} \), where each $\boldsymbol{X}^{(t)}$ represents the returns of $N$ assets at time $t$, and a user-specified confidence level $0 < \alpha < 1$ (e.g., 0.01 or 0.05), our goal is to construct a SCS, denoted by $\SCS$,  defined as a subset of $\Scal$ satisfying the coverage property 
\begin{equation}
\mathbb{P}(\Scal_0 \subseteq \SCS  ) \geq 1 - \alpha,
\end{equation}
asymptotically, as $T\to \infty$. This condition ensures that, with approximate probability of at least \( 1 - \alpha \) under repeated sampling, the true optimal selections \( \mathcal{S}_0 \) are included in the constructed set. In analogy to classical confidence intervals for parameter estimation, the SCS quantifies uncertainty in the portfolio selection process by identifying a set of EWPs whose empirical performance is statistically indistinguishable from $L_0$.

The SCS  supports two main tasks in portfolio selection. 1) It enables formal plausibility assessment of a given candidate portfolio \( \mathbf{s} \): by checking whether \( \mathbf{s} \in \SCS \), practitioners can formally test whether its performance is statistically indistinguishable from that of an unknown optimum. This allows for rigorous validation of alternative portfolio configurations. Moreover, comparing the asset composition of $\bfs$ with those of other plausible elements in $\SCS$ reveals robust inclusion patterns, highlighting complementary investment strategies and asset combinations that consistently contribute to strong performance.  2) The SCS quantifies the degree of selection uncertainty. Its size and structure reflect the informativeness of the data: a singleton SCS indicates a uniquely optimal allocation; a small but non-singleton points to manageable ambiguity; and a large SCS suggests high uncertainty, with many portfolios exhibiting comparable performance.

Overall, the SCS is a statistically grounded tool for assessing model uncertainty in portfolio selection. By identifying the full set of portfolios with performance statistically indistinguishable from a portfolio with minimal loss, it complements traditional EWPs optimization with formal inferential guarantees, helping decision-makers navigate the uncertainty inherent in financial allocation.

\subsection{Construction Based on Wald Test Screening} \label{sec:SCSconstruction}

To construct \( \SCS \), we aim to compare the population loss associated with each candidate selection $\bfs$ to that of any optimal selection $\bfs_0 \in \Scal_0$, using a Wald-type test.  To quantify the performance discrepancy between any selections \( \bfs, \bfs' \in \Scal \), we define the  performance differential as
$
\delta(\bfs; \bfs') := L(\mu_{\bfs}, \sigma^2_{\bfs}) - L(\mu_{\bfs'}, \sigma^2_{\bfs'})
$. To identify selections whose performance is statistically indistinguishable from the optimal, we test the hypotheses
\begin{align}\label{eq:H0}
H_0: \delta(\bfs) = 0 \quad \text{vs.} \quad H_1: \delta(\bfs) > 0,
\end{align}
where \( \delta(\bfs) := \delta(\bfs; \bfs_0) =  L(\mu_{\bfs}, \sigma^2_{\bfs}) - L_0 \), and \( \bfs_0 \in \Scal_0 \) is a fixed reference portfolio achieving the minimal  loss $L_0$. If $H_0$ is not rejected -- i.e., the data do not provide sufficient evidence to distinguish the candidate portfolio \( Y_{\bfs} \) from the reference \( Y_0 \) -- we include \( \bfs \) in the SCS.

Since the true optimal set \( \Scal_0 \) is unknown, we use an empirical proxy obtained by minimizing a sample-based loss over the feasible support space \( \Scal \). Given return observations \( \boldsymbol{X}^{(t)} = (X^{(t)}_1, \dots, X^{(t)}_N)^\top \in \mathbb{R}^N \), \( t = 1, \dots, T \),  corresponding to EWPs returns \( Y^{(t)}_{\boldsymbol{s}} \),  \( t = 1, \dots, T \),  we consider the sample mean and variance of portfolio returns for \( \boldsymbol{s} \in \Scal\),
\begin{equation}\label{eq:sample_moments}
\hat{\mu}_{\boldsymbol{s}} := \frac{1}{T} \sum_{t=1}^{T} Y^{(t)}_{\boldsymbol{s}}, \quad \hat{\sigma}^2_{\boldsymbol{s}} := \frac{1}{T - 1} \sum_{t=1}^{T} \left( Y^{(t)}_{\boldsymbol{s}} - \hat{\mu}_{\boldsymbol{s}} \right)^2,
\end{equation}
and define the empirical optimum as
\begin{equation}\label{eq:optimal_empirical}
\hat{\mathbf{s}}_0 := \underset{\mathbf{s} \in \Scal}{\arg\min}\; L(\hat\mu_{\mathbf{s}}, \hat\sigma^2_{\mathbf{s}}).
\end{equation}
Since the empirical loss is computed over a finite  set and portfolio moments are continuous functions of the data,  \( \hat{\mathbf{s}}_0 \) is unique with probability one under mild conditions. In contrast, the population-optimal set \( \Scal_0 \) may contain more than one element. Consequently, in regular settings, \( \hat{\mathbf{s}}_0 \) converges in probability to some element of \( \Scal_0 \), although not necessarily uniquely. This follows from the consistency of the sample means and variances used to evaluate the loss (see  proof of Proposition~\ref{prop:scs_coverage}), justifying the use of \( \hat{\mathbf{s}}_0 \) as a reference. However, since all elements in \( \Scal_0 \) attain the same minimal population loss $L_0$, the presence of multiple optima is not a concern and does not affect the validity of using \( \hat{\mathbf{s}}_0 \) as a reference, as any such limit point represents an equally optimal selection.

For a fixed \(  \alpha \), the SCS $\SCS$ is defined as
\begin{equation} \label{eq:scs}
\SCS := \left\{ \mathbf{s} \in \Scal : \; Z(\bfs; \hat{\bfs}_0) \leq q_{1-\alpha} \right\},
\end{equation}
where $Z(\bfs; \bfs')$ is the studentized differential for selections $\bfs, \bfs' \in \Scal$, defined by
\begin{equation}\label{eq:wald_statistic}
Z(\bfs; \bfs') := \dfrac{\hat{\delta}(\bfs; \bfs')}{\sqrt{\hat \tau(\bfs;\bfs')/T}} := \dfrac{L(\hat\mu_{\bfs}, \hat\sigma^2_{\bfs}) - L(\hat\mu_{\bfs'}, \hat\sigma^2_{\bfs'})}{\sqrt{\nabla \hat{\delta}(\bfs; \bfs')^\top \hat{\bfV}(\bfs; \bfs') \nabla \hat{\delta}(\bfs; \bfs') /T}},
\end{equation}
where $\hat{\delta}(\bfs; \bfs') = L(\hat\mu_{\bfs}, \hat\sigma^2_{\bfs}) - L(\hat\mu_{\bfs'}, \hat\sigma^2_{\bfs'})$ represents the empirical performance differential, $\hat \tau(\bfs; \bfs')^2$ is its estimated variance and $q_{1-\alpha}$ is an $(1-\alpha)$-quantile specified depending on the distribution of $Z(\bfs; \hat \bfs_0)$. In the denominator,  $\nabla \hat{\delta}(\bfs; \bfs')$ represents $4\times 1$  gradient  vector of the loss difference evaluated at the moment estimates $\hat{\boldsymbol{\theta}}(\bfs; \bfs') := 
(\hat\mu_{\bfs}, 
\hat\sigma^2_{\bfs}, 
\hat\mu_{\bfs'}, 
\hat\sigma^2_{\bfs'})^\top$, that is,
\begin{equation}
\nabla \hat \delta(\bfs; \bfs') :=
\left(
\dfrac{\partial}{\partial \hat \mu_{\bfs}}  \hat \delta(\bfs; \bfs'), 
\dfrac{\partial}{\partial \hat \sigma^2_{\bfs}}  \hat \delta(\bfs; \bfs')
, 
\dfrac{\partial}{\partial \hat \mu_{\bfs'}}  \hat \delta(\bfs; \bfs'), 
\dfrac{\partial}{\partial \hat \sigma^2_{\bfs'}}  \hat \delta(\bfs; \bfs')
\right)^\top,
\end{equation}
and \( \hat{\bfV}(\bfs; \bfs') \) is a consistent estimator of the asymptotic variance $\bfV(\bfs; \bfs')$ of the moment estimator $\hat \bftheta(\bfs;\bfs')$ defined in \eqref{eq:asy_variance}.

The selection confidence set \( \SCS \) is constructed by comparing each candidate \( \bfs \in \Scal \) to the empirical benchmark \( \hat{\bfs}_0 \). Since \( \hat{\bfs}_0 \) converges in probability to an element of the population-optimal set \( \Scal_0 \), the set \( \SCS \) includes, with high probability, all selections whose performance is statistically indistinguishable from the optimal loss \( L_0 \). Selections with higher loss than \( \hat{\bfs}_0 \) may still be included if their difference in performance is small compared to its uncertainty.  Rather than committing to the single best selection $\hat \bfs_0$, \( \SCS \) offers a robust alternative by retaining a number of well-performing selections, which is particularly useful when small performance differences do not justify distinct portfolio allocations.

The null distribution of \( Z(\bfs; \hat{\bfs}_0) \) can be specified in various ways. Here, we rely on a normal approximation justified by the central limit theorem. Particularly, for each $\bfs, \bfs' \in \Scal$, we assume that the empirical moments $\hat \bftheta(\bfs,\bfs')$ satisfy
\begin{equation}\label{eq:asy_distribution}
\sqrt{T} \left(  \hat{\bftheta}(\bfs; \bfs')  - \bftheta(\bfs; \bfs') \right) \overset{d}{\rightarrow} \mathcal{N}_4(\mathbf{0}, \bfV(\bfs; \bfs')),
\end{equation}
where $\bfV(\bfs; \bfs')$ is the $4\times 4$ matrix
\begin{equation} \label{eq:asy_variance}
\bfV(\bfs; \bfs') := 
\begin{pmatrix}
\bfV_{\bfs\bfs} := \operatorname{Cov}[(\hat\mu_{\bfs}, \hat\sigma^2_{\bfs})] & 
\bfV_{\bfs\bfs'} := \operatorname{Cov}[(\hat\mu_{\bfs}, \hat\sigma^2_{\bfs}), (\hat\mu_{\bfs'}, \hat\sigma^2_{\bfs'})] \\
\bfV_{\bfs'\bfs} := \operatorname{Cov}[(\hat\mu_{\bfs'}, \hat\sigma^2_{\bfs'}), (\hat\mu_{\bfs}, \hat\sigma^2_{\bfs})] & 
\bfV_{\bfs'\bfs'} := \operatorname{Cov}[(\hat\mu_{\bfs'}, \hat\sigma^2_{\bfs'})]
\end{pmatrix}.
\end{equation}
This relation holds under mild regularity conditions.
For example, when the data are assumed i.i.d., it is sufficient to have finite fourth moment $\mathbb{E}[Y^4_{\bfs}]<\infty$ for all $\bfs \in \Scal$. In the time series case, it is
sufficient to have finite $4+\epsilon$ moments, with $\epsilon$ being a small positive constant, together with an appropriate mixing condition; e.g., see \citep{andrews1991heteroskedasticity}.

If consistency of the sample moments in $\hat \bftheta(\bfs, \bfs')$ applies uniformly for $\bfs, \bfs' \in \Scal$, the empirical loss converges to the population loss, and the event $\hat \bfs_0 \in \Scal_0$ occurs with probability tending to one. Therefore, the test statistic \( Z(\bfs; \hat{\bfs} _0) \) under the null hypothesis $H_0: \delta(\bfs) = 0$ converges in distribution to a standard normal distribution by the Delta method, that is $Z(\bfs; \hat{\bfs}) \overset{d}{\rightarrow} \mathcal{N}(0, 1)$ .  Therefore, to retain $1-\alpha$  coverage asymptotically, the quantile $q_{1-\alpha}$ in \eqref{eq:scs} is set as the $(1-\alpha)$ quantile of the standard normal distribution.

While the above test serves as our default screening tool due to its flexibility in accommodating a wide class of loss functions and its reliability in large samples, the SCS framework is compatible with any alternative test specification for the null $H_0: \delta(\bfs)=0$. Examples include the finite-sample Sharpe ratio test of \citet{Me.03} and the robust bootstrap procedure of \citet{LW.08}. Extensions to variance (e.g., \cite{LW.11}, implemented in \cite{Din.21}) and smooth functionals of higher-order moments (e.g., \cite{LW.18} implemented in \cite{Lei.23}) further broaden the applicability. While these tests are typically used for ad hoc pairwise comparisons, our framework incorporates them to construct an SCS that covers the population-optimal selection with the prescribed confidence level.

\subsection{Screening under i.i.d. sampling and loss-specific criteria}

If portfolio returns \( Y^{(1)}_{\bfs}, \dots, Y^{(T)}_{\bfs} \) form an i.i.d. sequence for all \( \bfs \in \Scal \), then the sample moments associated with any pair of selections \( (\bfs, \bfs') \) admit the limiting distribution in \eqref{eq:asy_distribution}, with the following blocks for $\bfV(\bfs; \bfs')$:
\[
\bfV_{\bfs\bfs} =  
\begin{pmatrix}
\sigma_{\bfs}^2 & \mu_{3,\bfs} \\
\mu_{3,\bfs} & \mu_{4,\bfs} - \sigma_{\bfs}^4
\end{pmatrix}, 
\bfV_{\bfs\bfs'} =  
\begin{pmatrix}
\sigma_{\bfs\bfs'} & \mu_{3,\bfs\bfs'} \\
\mu_{3,\bfs'\bfs} & \mu_{4,\bfs\bfs'} - \sigma_{\bfs\bfs'}^2
\end{pmatrix}, 
\bfV_{\bfs'\bfs'} =  
\begin{pmatrix}
\sigma_{\bfs'}^2 & \mu_{3,\bfs'} \\
\mu_{3,\bfs'} & \mu_{4,\bfs'} - \sigma_{\bfs'}^4
\end{pmatrix}
\]
and $\bfV_{\bfs'\bfs} = \bfV_{\bfs\bfs'}^\top$. This  structure captures both marginal and joint variation up to the fourth moment, which is necessary to account for tail risk and asymmetries in performance differentials. Specifically, the third-order moments include the marginal skewness terms \( \mu_{3,\bfs} = \mathbb{E}[(Y_{\bfs} - \mu_{\bfs})^3] \) and \( \mu_{3,\bfs'} = \mathbb{E}[(Y_{\bfs'} - \mu_{\bfs'})^3] \), as well as the mixed skewness terms \( \mu_{3,\bfs\bfs'} = \mathbb{E}[(Y_{\bfs} - \mu_{\bfs})(Y_{\bfs'} - \mu_{\bfs'})^2] \) and \( \mu_{3,\bfs'\bfs} = \mathbb{E}[(Y_{\bfs'} - \mu_{\bfs'})(Y_{\bfs} - \mu_{\bfs})^2] \). The fourth-order moments include the marginal kurtoses \( \mu_{4,\bfs} = \mathbb{E}[(Y_{\bfs} - \mu_{\bfs})^4] \), \( \mu_{4,\bfs'} = \mathbb{E}[(Y_{\bfs'} - \mu_{\bfs'})^4] \), and the joint kurtotic term \( \mu_{4,\bfs\bfs'} = \mathbb{E}[(Y_{\bfs} - \mu_{\bfs})^2 (Y_{\bfs'} - \mu_{\bfs'})^2] \). In this case,  \( \bfV (\bfs; \bfs')\)  is estimated by substituting the sample means and variances into the analytical form derived above.

The special case where \( Y^{(t)}_{\bfs} \) is normally distributed serves as a benchmark for EWP selection under idealized conditions. Under normality, all third-order moments vanish; namely,  \( \mu_{3,\bfs} = \mu_{3,\bfs'} = \mu_{3,\bfs\bfs'} = \mu_{3,\bfs'\bfs} = 0 \). Moreover, fourth-order moments simplify as: \( \mu_{4,\bfs} = 3\sigma_{\bfs}^4 \), \( \mu_{4,\bfs'} = 3\sigma_{\bfs'}^4 \), and \( \mu_{4,\bfs\bfs'} = \sigma_{\bfs\bfs'}^2 + 2\sigma_{\bfs}^2\sigma_{\bfs'}^2 \), thus leading to simple expressions for screening statistics under common loss functions \( L(\mu, \sigma^2) \). For example,  for the mean-variance loss   $
L(\mu, \sigma^2) = -\gamma \mu + \sigma^2$, $\gamma > 0$, 
where \( \gamma \) denotes the risk aversion coefficient, the SCS is constructed by testing 
$
H_0: \delta(\bfs; \bfs_0) = \gamma(\mu_{\bfs_0} - \mu_{\bfs}) + \sigma^2_{\bfs} - \sigma^2_{\bfs_0} = 0$. This leads to
\begin{equation}\label{eq:SCS_mean_variance}
\SCS = \left\{ \bfs \in \Scal: Z(\bfs; \hat{\bfs}_0) = \dfrac{
\sqrt{T}\left[ \gamma(\hat{\mu}_{\hat{\bfs}_0} - \hat{\mu}_{\bfs}) + \hat{\sigma}^2_{\bfs} - \hat{\sigma}^2_{\hat{\bfs}_0} \right]
}{
\sqrt{ \gamma^2\left( \hat{\sigma}^2_{\bfs} - 2 \hat{\sigma}_{\bfs \hat{\bfs}_0} + \hat{\sigma}^2_{\hat{\bfs}_0} \right)
+ 2\left( \hat{\sigma}^4_{\bfs} - 2 \hat{\sigma}^2_{\bfs \hat{\bfs}_0} + \hat{\sigma}^4_{\hat{\bfs}_0} \right) }
} \leq q_{1-\alpha}
\right\}. \notag
\end{equation}
For the Sharpe ratio loss
$
L(\mu, \sigma^2) = - \mu/\sigma$,  the SCS is constructed by testing  $
H_0: \delta(\bfs; \bfs_0) = {\mu_{\bfs_0}}/{\sigma_{\bfs_0}} - {\mu_{\bfs}}/{\sigma_{\bfs}} = 0$. Thus, we obtain
\begin{equation}\label{eq:SCS_sharpe}
\SCS = \left\{ \bfs \in \Scal: 
Z(\bfs; \hat{\bfs}_0) = 
\dfrac{
\sqrt{T}(\hat r_{\bfs} - \hat r_{\hat{\bfs}_0})
}{
\sqrt{ 2(1 - \hat{\rho}_{\bfs \hat{\bfs}_0}) + \tfrac{1}{2}(\hat r_{\bfs}^2 + \hat r_{\hat{\bfs}_0}^2) - \hat{\rho}_{\bfs \hat{\bfs}_0}^2 \hat r_{\bfs} \hat r_{\hat{\bfs}_0} }
} \leq q_{1-\alpha}
\right\},
\end{equation}
where \( \hat r_{\bfs} = \hat \mu_{\bfs} / \hat \sigma_{\bfs} \) and \( \hat \rho_{\bfs \hat{\bfs}_0} = \hat \sigma_{\bfs \hat{\bfs}_0} / (\hat \sigma_{\bfs} \hat \sigma_{\hat{\bfs}_0}) \).

\section{Asymptotic Validity and Size} \label{sec:properties}

In this section, we study the asymptotic  coverage probability for the set of optimal selections \( \mathcal{S}_0 \) and the SCS size 
 as \( T \to \infty \), assuming a fixed number of  assets \( N < \infty \). To establish the asymptotic properties of the SCS, we proceed under high-level conditions that ensure valid inference without requiring restrictive model assumptions.  For all \( \bfs, \bfs' \in \Scal \) we assume:
\begin{itemize}
\item[(A1)] The vector of estimators \( (\hat{\mu}_{\bfs}, \hat{\sigma}^2_{\bfs}, \hat{\mu}_{\bfs'}, \hat{\sigma}^2_{\bfs'}) \) jointly satisfies the central limit theorem as in Equation~\eqref{eq:asy_distribution}, with asymptotic covariance matrix \( \bfV(\bfs, \bfs') \).
\item[(A2)] Each element of the estimated covariance matrix $\hat \bfV(\bfs, \bfs')$ converges in probability to its population counterpart $
\hat{\bfV}_{jk}(\bfs, \bfs') \overset{p}{\to} \bfV_{jk}(\bfs, \bfs')$, for all $j,k \in \{1,2,3,4\}$.
\end{itemize}
Next, we show that the SCS defined in (\ref{eq:scs}) contains the true set of optimal selections \( \mathcal{S}_0 \) with probability of at least \( 1 - \alpha \) in large samples.
\begin{proposition}[Asymptotic coverage of the optimal set]
\label{prop:scs_coverage}
Under Assumptions (A1)--(A2),  the Selection Confidence Set \( \SCS \) defined in~\eqref{eq:scs} satisfies the asymptotic coverage property:
\[
\lim_{T \to \infty} \mathbb{P}\left( \mathcal{S}_0 \subseteq \SCS \right) \geq 1 - \alpha.
\]
\end{proposition}

The result relies on two key aspects implied by Assumptions (A1) and (A2): (i) convergence in probability of sample moments \( (\hat{\mu}_{\bfs}, \hat{\sigma}^2_{\bfs}) \) to their population counterparts, ensuring that \( \hat{\bfs}_0\) converges to \( \bfs_0 \in \mathcal{S}_0 \); and (ii) convergence in distribution of the test statistic \( Z(\bfs, \hat{\bfs}_0) \) under the null, ensuring that our test has asymptotic size $\alpha$. As already mentioned, these hold under general dependence and heteroskedasticity in asset returns, requiring only that sample moments of portfolio returns admit a joint asymptotic normal distribution.
In finite samples, $\SCS$ may include many suboptimal portfolios, especially when loss differentials are small or variance estimates are noisy, reducing the power of pairwise tests. A large SCS is not a flaw but a reflection of data uncertainty, necessary for valid coverage. Hence, characterizing its size and structure is crucial for assessing selection ambiguity. To this end, we introduce the standardized loss differential
\begin{equation}
\gamma(\bfs) := \dfrac{\delta(\bfs)}{\tau(\bfs)} = \dfrac{L(\mu_{\mathbf{s}}, \sigma^2_{\mathbf{s}}) - L(\mu_0, \sigma^2_0)}{\sqrt{\nabla \delta(\bfs,\bfs_0)^\top \mathbf{V}(\bfs ,\bfs_0) \nabla \delta(\bfs,\bfs_0)}} ,
\end{equation}
 where $\nabla \delta(\bfs,\bfs_0)$ is the gradient  of $\delta(\bfs,\bfs_0)$ with respect to $(\mu_{\bfs}, \sigma^2_{\bfs}, \mu_{\bfs_0}, \sigma^2_{\bfs_0} )^\top$. This ratio captures the signal strength relative to estimation uncertainty and governs the probability that a non-optimal support is erroneously included in the SCS. The next proposition makes this relationship precise by characterizing the limiting expected size of the SCS.

\begin{proposition}[Asymptotic expected size of the SCS]
\label{prop:asymptotic_scs_size}
Under assumptions (A1)--(A2), the expected size of the Selection Confidence Set \( \SCS \) satisfies
\begin{equation}\label{eq:expected_size}
\mathbb{E}[|\SCS|] = |\mathcal{S}_0|(1 - \alpha) + \sum_{\bfs \in \mathcal{S} \setminus \mathcal{S}_0}
\Phi\left( q_{1 - \alpha} -  \sqrt{T}  \gamma(\bfs)  \right) + o(1)
\end{equation}
where $\Phi(\cdot)$ is the cdf of the standard normal distribution, and $q_{1-\alpha}$ is its $(1-\alpha)$ quantile.
\end{proposition}

Proposition~\ref{prop:asymptotic_scs_size} shows the relationship between  the inclusion probability of a suboptimal asset selection \( \bfs \notin \mathcal{S}_0 \) and its standardized differential \( \gamma(\bfs) \). If \( \sqrt{T} \gamma(\bfs) \to \infty \), then the inclusion probability \( \Phi(z_{1-\alpha} - \sqrt{T} \gamma(\bfs)) \) decays exponentially, and $\bfs$ is eventually excluded from \( \SCS \). Thus, the cardinality of the SCS is driven by the number of suboptimal supports for which \( \gamma(\bfs) = o(T^{-1/2}) \), as these remain empirically indistinguishable from optimal ones. In this sense, the SCS quantifies the set of portfolios that are statistically non-separable from the best-performing configurations, and its size reflects the inherent difficulty of model selection under limited information.

To further elaborate, let $\gamma_{\text{min}} := \min_{\mathbf{s} \notin \mathcal{S}_0}  {\delta(\bfs)}/{\tau(\bfs)}$
denote the minimal standardized signal among selections outside the optimal set.  Since the function $
x \mapsto   \Phi( q_{1-\alpha} - \sqrt{T}x )$
is monotonically decreasing in \( x \),  from (\ref{eq:expected_size}), we immediately obtain the approximate bounds
\begin{equation}\label{eq:scs_size_bound}
|\mathcal{S}_0|(1 - \alpha) + o(1)\leq \mathbb{E}[|\SCS|]  
\leq |\mathcal{S}_0|(1 - \alpha) + \left(2^N - |\mathcal{S}_0|-1\right) \Phi\left(z_{1-\alpha} - \sqrt{T} \gamma_{\text{min}}\right) +  o(1).
\end{equation}
If \( \gamma_{\min} \) is bounded away from zero -- i.e., all suboptimal models are well-separated from the optimal set -- then the expected SCS size concentrates around \( |\mathcal{S}_0|(1 - \alpha) \), reflecting the expected retention of optimal supports. In contrast, when \( \gamma_{\min} \) is small, the inclusion probability \( \Phi(z_{1-\alpha} - \sqrt{T} \gamma_{\min}) \) remains large, and many suboptimal models may enter \( \mathcal{S}_\alpha \). 

\section{Post-selection Metrics} 
\label{sec:metrics}

We introduce descriptive statistics derived from the SCS to summarize uncertainty in portfolio selection. These include global measures of multiplicity, performance variability, and asset-level inclusion patterns.

{\it Global selection uncertainty.} Basic summaries are the  cardinality \( |\SCS| \), the  relative cardinality  \( |\SCS| /|\Scal| \) and its scale-adjusted version,  the Relative Multiplicity Index (RMI):
\begin{equation} \label{RMI}
\mathrm{RMI} := 1- \dfrac{\log(|\SCS|)}{\log(|\Scal|)},
\end{equation}
which ranges from 0 (maximal uncertainty) to 1 (unique selection). The logarithmic transformation accounts for the natural scale of growth in \( |\Scal| \), which is typically exponential in the number of assets. In addition, to quantify the empirical range of acceptable losses we consider the observed loss interval $(\hat{L}_0, \hat L_{\text{max}})$, where $\hat{L}_0 = L(\hat{\mu}_{\hat{\bfs}_0}, \hat{\sigma}^2_{\hat{\bfs}_0})$ and $\hat L_{\text{max}} := \max_{\bfs \in \mathcal{S}_\alpha} L(\hat{\mu}_{\bfs}, \hat{\sigma}^2_{\bfs})$ 
and the associated performance spread $\hat \delta_{\alpha} := 
\hat L_{\text{max}} - \hat{L}_0$. These are especially relevant in applied risk-sensitive settings; for example a  manager may accept a candidate allocation \( \bfs \in \mathcal{S}_\alpha \) only if this metrics falls below a tolerable threshold.

\textit{Lower-boundary (LB) portfolios.} The lower boundary \( \LB \subseteq  \mathcal{S}_\alpha \) contains the most parsimonious selections in the SCS: $
\LB := \left\{ \bfs \in \SCS : \nexists\, \bfs' \in \SCS \text{ with } \text{supp}(\bfs') \subset \text{supp}(\bfs) \right\}$.
These are minimal configurations that cannot be further reduced without losing information, revealing which assets are essential for achieving acceptable risk-return trade-offs. As such, they provide a data-driven foundation for lean investment strategies -- especially valuable when transaction costs, regulatory limits, or interpretability concerns favor compact allocations.

\textit{Asset inclusion and co-inclusion importance.} The inclusion importance of asset \( j \in \{1, \dots, N\} \) measures how often it appears across portfolios in the SCS:
\[
\mathrm{II}_\alpha(j) := \frac{1}{|\SCS|} \sum_{\bfs \in \SCS} s_j, 
\]
with value close to 1 indicates that asset \( j \) appears in nearly all selected portfolios, while a value near 0 indicates rare or no inclusion.  To quantify how often assets \( i \) and \( j \) appear together in the SCS, we define their co-inclusion importance as
\[
\mathrm{CII}_\alpha(i, j) := 
 \dfrac{ \sum_{\bfs \in \SCS} s_i s_j }{ \sum_{\bfs \in \SCS} s_i + \sum_{\bfs \in \SCS} s_j - \sum_{\bfs \in \SCS} s_i s_j },
\]
if the denominator is strictly positive and $\mathrm{CII}_\alpha(i, j) := 1$ otherwise. This index measures how frequently assets \( i \) and \( j \) co-occur relative to their marginal inclusion frequencies. High values indicate complementarity, while low values reflect mutual exclusivity. To visualize these patterns, one can construct an undirected asset graph where nodes represent assets and edge weights proportional to \( \mathrm{CII}_\alpha(i, j) \), revealing clusters, substitution patterns, and diversification structures.

\section{Numerical Experiments}
\label{sec:Monte_Carlo}

\subsection{Simulation Setup} \label{sec:simulation_setup}

 Asset returns \( \boldsymbol{X} \in \mathbb{R}^N \) are drawn from a multivariate normal distribution with mean \( \boldsymbol{\eta} \in \mathbb{R}^N \) and covariance \( \boldsymbol{\Sigma} = \mathbf{D} \mathbf{R} \mathbf{D} \), where \( \mathbf{R} \in \mathbb{R}^{N \times N} \) is the correlation matrix and \( \mathbf{D} = \mathrm{diag}(\sqrt{\Sigma_{11}}, \dots, \sqrt{\Sigma_{NN}}) \) is a diagonal matrix of marginal standard deviations. To reflect realistic market conditions with a weak positive risk–return association, we set \( \eta_j = -0.002 + \Sigma_{jj}/10 + \varepsilon_j \), where marginal variances \( \Sigma_{jj} \) are drawn independently from a uniform distribution on \([0.01, 0.03] \), and \( \varepsilon_j \sim \mathcal{N}(0, 0.02) \) for \( j = 1, \dots, N \). The correlation matrix \( \mathbf{R} \) is set according to the following asset dependence models. 
 
 \noindent \textit{Model 1) Graph structure.} We generate an inverse covariance matrix \( \boldsymbol{\Omega}^{-1} \) based on a scale-free graph structure, reflecting empirically observed dependencies in asset returns \citep{K.10}. Using the \texttt{huge.generator} function from the \texttt{R} package \texttt{huge} \citep{huge}, we construct a sparse adjacency matrix \( \boldsymbol{\Theta} \), where off-diagonal entries of \( \boldsymbol{\Omega} \) corresponding to edges are set to a constant \( v > 0 \).  To ensure positive definiteness,  diagonal entries are adjusted as \( \boldsymbol{\Omega}_{jj} = |e| + 0.2 \), with \( e \) being the smallest eigenvalue of \( v \boldsymbol{\Theta} \). We consider \(v \in \{0.2,1\}\) to account for mild and strong conditional dependence. The resulting precision matrix is inverted and rescaled to obtain the correlation matrix \( \mathbf{R} \). 
 
 \noindent{\it Model 2) Exchangeable structure}. We set $\{\mathbf{R} \}_{jk} = \rho$ for $j\neq k$ with $\rho \in \{0.25, 0.75 \}$ corresponding to mild and strong correlation.

For various values of \( T \) and \( N \), we compute Monte Carlo estimates of the expected SCS cardinality \( \kappa_\alpha := \mathbb{E}[|\SCS|] \), that of its lower boundary \( \underline{\kappa}_\alpha := \mathbb{E}[|\LB|] \), and the coverage probability \( p_{\alpha} := \mathbb{P}(\mathbf{s}_0 \in \SCS) \), based on 300 simulation runs. For each run, we construct SCSs at confidence levels 90\%, 95\%, and 99\%, using the Sharpe ratio loss (\( -\mu/\sigma \)), mean-variance loss (\( -0.5\,\mu + \sigma^2 \)), and the expected shortfall (ES) loss (\( -\mu + \sigma\,\phi(z_{0.1})/0.1 \)), where $\phi(\cdot)$ is the pdf of the standard normal distribution. In all scenarios, the true and empirical optimal EWPs are identified via exhaustive search.

\subsection{Results}

In Table~\ref{table:model1_v1}, we report Monte Carlo estimates of \( \kappa_\alpha \), \( \underline{\kappa}_\alpha \), and  \( p_\alpha \) for Model 1 under strong conditional dependence (\( v = 1 \)) with fixed universe size (\( N = 10 \)). Table~\ref{table:model2_rho075} shows analogous results for Model 2, under strong correlation (\(\rho = 0.75 \)). Analogous results for Model 1 (with \( v = 0.2 \)) and Model 2 (with \( \rho = 0.25 \)) are reported in Table S1 and S2 in the Supplementary Materials.  Across all settings, the size of the SCS decreases rapidly with \( T \). As expected, higher confidence levels \( 1 - \alpha \) lead to more inclusive screening, yielding larger SCSs and improved coverage. For sufficiently large \( T \), the coverage  approaches or exceeds the nominal level \( 1 - \alpha \), with the only exception being expected shortfall (ES) loss with $v=0.2$, which requires $T>1000$ to reach the nominal coverage, confirming the asymptotic validity and size properties established in Section~\ref{sec:properties}. For smaller \( T \), coverage often falls below the nominal level, reflecting the limitations of asymptotic approximations and increased test statistic variance in finite samples.

\begin{table}[ht]
\centering
\begin{tabular}{c c ccc ccc ccc}
\toprule
\multicolumn{1}{c}{$L(\mu,\sigma^2)$} & \multicolumn{1}{c}{$T$}
   & \multicolumn{3}{c}{$1-\alpha=0.90$}
   & \multicolumn{3}{c}{$1-\alpha=0.95$}
   & \multicolumn{3}{c}{$1-\alpha=0.99$} \\
\cmidrule(lr){3-5} \cmidrule(lr){6-8} \cmidrule(l){9-11}
 &  & $\kappa_\alpha$ & $p_\alpha\%$ & $\underline{\kappa}_\alpha$
   & $\kappa_\alpha$ & $p_\alpha\%$ & $\underline{\kappa}_\alpha$
   & $\kappa_\alpha$ & $p_\alpha\%$ & $\underline{\kappa}_\alpha$ \\
\midrule
$-\mu/\sigma$
   & 100 & \makecell{125.7\\(7.0)} & \makecell{71.3\\(2.6)} & \makecell{ 3.4\\(0.1)}
         & \makecell{294.5\\(13.0)} & \makecell{84.7\\(2.1)} & \makecell{ 5.1\\(0.1)}
         & \makecell{630.9\\(14.8)} & \makecell{97.3\\(1.0)} & \makecell{ 7.6\\(0.1)} \\
 & 250 & \makecell{ 67.5\\(4.7)} & \makecell{82.3\\(2.2)} & \makecell{ 2.8\\(0.1)}
         & \makecell{143.2\\(6.8)}  & \makecell{91.7\\(1.6)} & \makecell{ 4.1\\(0.1)}
         & \makecell{418.8\\(13.2)} & \makecell{99.3\\(0.6)} & \makecell{ 6.4\\(0.1)} \\
 &1000 & \makecell{ 13.1\\(0.7)} & \makecell{89.0\\(1.8)} & \makecell{ 1.6\\(0.1)}
         & \makecell{ 25.4\\(1.4)}  & \makecell{98.0\\(0.8)} & \makecell{ 2.2\\(0.1)}
         & \makecell{ 81.2\\(3.8)}  & \makecell{99.3\\(0.6)} & \makecell{ 3.6\\(0.1)} \\
\cmidrule{1-11}
$-0.5\,\mu+\sigma^2$
   & 100 & \makecell{290.4\\(19.5)} & \makecell{71.0\\(2.6)} & \makecell{ 4.4\\(0.1)}
         & \makecell{523.6\\(21.5)} & \makecell{87.7\\(2.0)} & \makecell{ 5.9\\(0.1)}
         & \makecell{864.5\\(16.5)} & \makecell{96.3\\(1.2)} & \makecell{ 8.3\\(0.1)} \\
 & 250 & \makecell{167.6\\(12.3)} & \makecell{84.3\\(2.3)} & \makecell{ 3.6\\(0.1)}
         & \makecell{364.0\\(19.1)} & \makecell{91.0\\(1.5)} & \makecell{ 4.7\\(0.1)}
         & \makecell{727.9\\(18.8)} & \makecell{100.0\\(0.0)}& \makecell{ 6.9\\(0.1)} \\
 &1000 & \makecell{ 19.7\\(1.6)} & \makecell{92.3\\(1.4)} & \makecell{ 2.2\\(0.1)}
         & \makecell{ 56.0\\(5.3)} & \makecell{96.3\\(0.8)} & \makecell{ 2.7\\(0.1)}
         & \makecell{205.5\\(13.4)} & \makecell{100.0\\(0.0)}& \makecell{ 3.9\\(0.1)} \\
\cmidrule{1-11}
$-\mu + \sigma\,\frac{\phi(z_{0.1})}{0.1}$
   & 100 & \makecell{  8.9\\(0.3)} & \makecell{71.0\\(2.6)} & \makecell{ 1.6\\(0.1)}
         & \makecell{ 14.6\\(0.5)} & \makecell{81.0\\(2.3)} & \makecell{ 2.2\\(0.1)}
         & \makecell{ 33.1\\(1.1)} & \makecell{90.7\\(1.7)} & \makecell{ 3.8\\(0.2)} \\
 & 250 & \makecell{  5.9\\(0.2)} & \makecell{80.0\\(2.3)} & \makecell{ 1.3\\(0.0)}
         & \makecell{  9.2\\(0.3)} & \makecell{88.3\\(1.9)} & \makecell{ 1.3\\(0.0)}
         & \makecell{ 15.2\\(0.4)} & \makecell{96.3\\(1.1)} & \makecell{ 1.6\\(0.1)} \\
 &1000 & \makecell{  3.0\\(0.1)} & \makecell{95.3\\(1.2)} & \makecell{ 1.1\\(0.0)}
         & \makecell{  4.4\\(0.1)} & \makecell{97.7\\(0.9)} & \makecell{ 1.2\\(0.0)}
         & \makecell{  6.6\\(0.2)} & \makecell{100.0\\(0.0)} & \makecell{ 1.2\\(0.0)} \\

\bottomrule
\end{tabular}
\caption{Monte Carlo estimates of SCS size \( \kappa_\alpha \), coverage probability \( p_{\alpha} \) (in \%), and lower boundary size \( \underline{\kappa}_\alpha \) under the Sharpe ratio loss (\( -\mu/\sigma \)), mean-variance loss (\( -0.5\,\mu + \sigma^2 \)), and expected shortfall loss (\( -\mu + \sigma\,\phi(z_{0.1})/0.1 \)). Data are generated from Model~1 with \( v = 1 \)  and \( N = 10 \). Monte Carlo standard errors are shown in parentheses.}
\label{table:model1_v1}
\end{table}

\begin{table}[ht]
\centering
\begin{tabular}{c c ccc ccc ccc}
\toprule
\multicolumn{1}{c}{$L(\mu,\sigma^2)$} & \multicolumn{1}{c}{$T$}
   & \multicolumn{3}{c}{$1-\alpha=0.90$}
   & \multicolumn{3}{c}{$1-\alpha=0.95$}
   & \multicolumn{3}{c}{$1-\alpha=0.99$} \\
\cmidrule(lr){3-5} \cmidrule(lr){6-8} \cmidrule(l){9-11}
 &  & $\kappa_\alpha$ & $p_\alpha\%$ & $\underline{\kappa}_\alpha$
   & $\kappa_\alpha$ & $p_\alpha\%$ & $\underline{\kappa}_\alpha$
   & $\kappa_\alpha$ & $p_\alpha\%$ & $\underline{\kappa}_\alpha$ \\
\midrule
$-\mu/\sigma$
   & 100 & \makecell{74.9\\(7.2)} & \makecell{90.7\\(1.7)} & \makecell{3.6\\(0.1)}
         & \makecell{196.4\\(14.2)} & \makecell{95.7\\(1.2)} & \makecell{4.8\\(0.1)}
         & \makecell{557.3\\(20.5)} & \makecell{98.7\\(0.7)} & \makecell{6.9\\(0.1)} \\
 & 250 & \makecell{17.8\\(2.3)} & \makecell{95.3\\(1.2)} & \makecell{2.6\\(0.1)}
         & \makecell{60.4\\(5.8)} & \makecell{98.0\\(0.8)} & \makecell{3.6\\(0.1)}
         & \makecell{223.3\\(13.2)} & \makecell{99.3\\(0.5)} & \makecell{5.3\\(0.1)} \\
 &1000 & \makecell{3.3\\(0.1)} & \makecell{97.0\\(1.0)} & \makecell{1.8\\(0.0)}
         & \makecell{4.7\\(0.2)} & \makecell{99.7\\(0.3)} & \makecell{2.1\\(0.0)}
         & \makecell{10.8\\(0.8)} & \makecell{100.0\\(0.0)} & \makecell{2.5\\(0.0)} \\
\cmidrule{1-11}
$-0.5\,\mu+\sigma^2$
   & 100 & \makecell{116.4\\(11.1)} & \makecell{90.7\\(1.7)} & \makecell{4.0\\(0.1)}
         & \makecell{306.6\\(18.8)} & \makecell{96.0\\(1.1)} & \makecell{5.5\\(0.1)}
         & \makecell{645.2\\(20.7)} & \makecell{99.7\\(0.3)} & \makecell{7.8\\(0.1)} \\
 & 250 & \makecell{32.4\\(3.5)} & \makecell{94.0\\(1.4)} & \makecell{2.9\\(0.1)}
         & \makecell{73.6\\(7.5)} & \makecell{98.3\\(0.7)} & \makecell{3.5\\(0.1)}
         & \makecell{321.9\\(17.2)} & \makecell{99.7\\(0.3)} & \makecell{5.7\\(0.1)} \\
 &1000 & \makecell{3.4\\(0.1)} & \makecell{97.3\\(0.9)} & \makecell{1.8\\(0.0)}
         & \makecell{5.8\\(0.4)} & \makecell{99.3\\(0.5)} & \makecell{2.0\\(0.0)}
         & \makecell{14.5\\(1.2)} & \makecell{100.0\\(0.0)} & \makecell{2.5\\(0.0)} \\
\cmidrule{1-11}
$-\mu + \sigma\,\frac{\phi(z_{0.1})}{0.1}$
   & 100 & \makecell{2.2\\(0.1)} & \makecell{91.0\\(1.7)} & \makecell{1.0\\(0.0)}
         & \makecell{2.6\\(0.1)} & \makecell{97.3\\(0.9)} & \makecell{1.1\\(0.0)}
         & \makecell{4.1\\(0.1)} & \makecell{98.3\\(0.7)} & \makecell{1.3\\(0.0)} \\
 & 250 & \makecell{1.8\\(0.1)} & \makecell{97.0\\(1.0)} & \makecell{1.0\\(0.0)}
         & \makecell{2.1\\(0.1)} & \makecell{99.0\\(0.6)} & \makecell{1.0\\(0.0)}
         & \makecell{2.8\\(0.1)} & \makecell{99.7\\(0.3)} & \makecell{1.0\\(0.0)} \\
 &1000 & \makecell{1.2\\(0.0)} & \makecell{99.7\\(0.3)} & \makecell{1.0\\(0.0)}
         & \makecell{1.4\\(0.0)} & \makecell{100.0\\(0.0)} & \makecell{1.0\\(0.0)}
         & \makecell{1.6\\(0.0)} & \makecell{100.0\\(0.0)} & \makecell{1.0\\(0.0)} \\
\bottomrule
\end{tabular}
\caption{Monte Carlo estimates of SCS size \( \kappa_\alpha \), coverage probability \( p_{\alpha} \) (in \%), and lower boundary size \( \underline{\kappa}_\alpha \) under the Sharpe ratio loss (\( -\mu/\sigma \)), mean-variance loss (\( -0.5\,\mu + \sigma^2 \)), and expected shortfall loss (\( -\mu + \sigma\,\phi(z_{0.1})/0.1\)). Data are generated from Model~2 with \( \rho = 0.75 \) and \( N = 10 \). Monte Carlo standard errors are shown in parentheses.}
\label{table:model2_rho075}
\end{table}

The size of the lower boundary subset $\LB$ is substantially smaller than that of the SCS $\SCS$, often comprising fewer than ten portfolio selections. Its constituent selections typically correspond to portfolios concentrated on a small number of assets. These sparse combinations have higher variance in estimated performance and are thus more likely to attain extreme empirical performance near the decision boundary. Similarly to the full SCS, the size of \( \LB \) decreases with \( T \) and both sets eventually concentrate on the true optimal set \( \mathcal{S}_0 \) as \( T \to \infty \).

We observe that both the asset dependence structure and loss function affect selection uncertainty, as reflected in the size of the SCS. For example, in Model~1, the stronger conditional dependence with \( v =1\) reduces the SCS size under the mean-variance loss compared to \( v =0.2\). Stronger dependence increases portfolio variances \( \sigma_{\bfs} \), amplifying the loss differentials \( \hat{\delta}(\bfs) \) across selections. As this signal gain outweighs the rise in estimation noise, the power of the screening test improves, resulting in smaller SCSs. We find that the choice of loss function leads to heterogeneous SCS sizes, primarily due to different sensitivity to estimation error. For example, the mean–variance loss is more sensitive than the ES loss, resulting in larger confidence sets. By contrast, the ES loss yields more stable inference and requires smaller samples to achieve comparable precision.

\begin{figure}[ht]
\centering

\begin{subfigure}[t]{0.32\textwidth}
    \centering
    $~~~~~-\mu/\sigma$\\[-0.3em]
    \includegraphics[width=\linewidth]{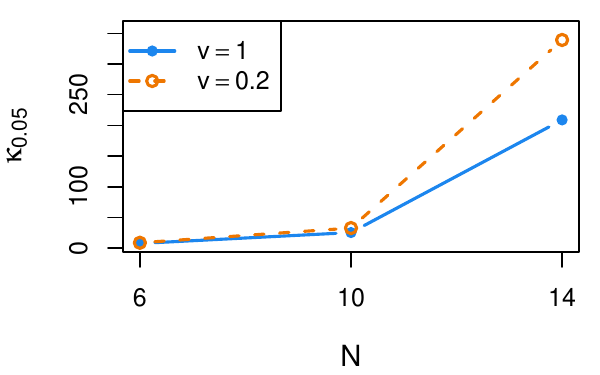}\\
    \includegraphics[width=\linewidth]{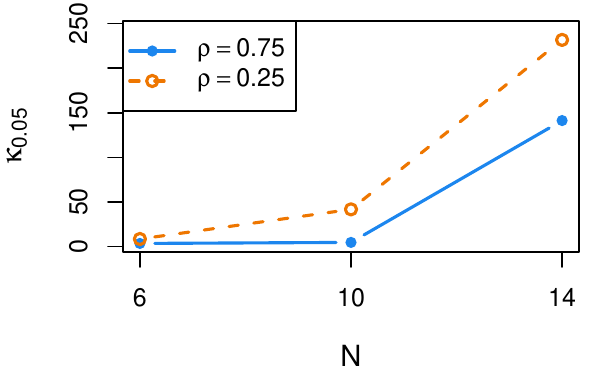}
\end{subfigure}
\hfill
\begin{subfigure}[t]{0.32\textwidth}
    \centering
    $~~~~~-0.5\,\mu+\sigma^2$\\[-0.3em]
    \includegraphics[width=\linewidth]{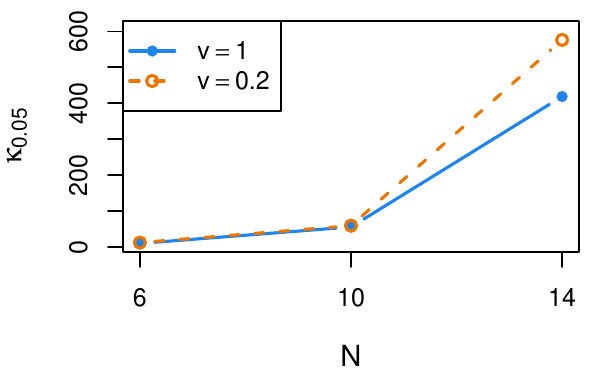}\\
    \includegraphics[width=\linewidth]{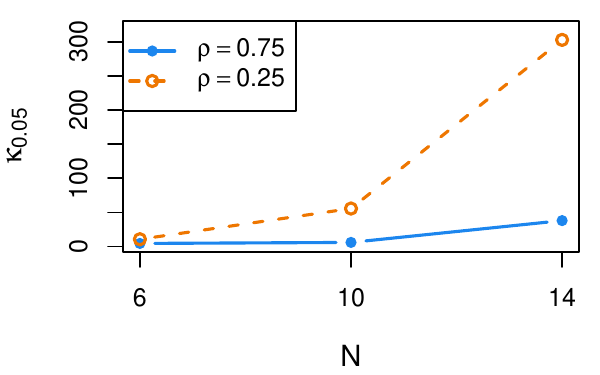}
\end{subfigure}
\hfill
\begin{subfigure}[t]{0.32\textwidth}
    \centering
    $~~~~~-\mu + \sigma\,\frac{\phi(z_{0.1})}{0.1}$\\[-0.3em]
    \includegraphics[width=\linewidth]{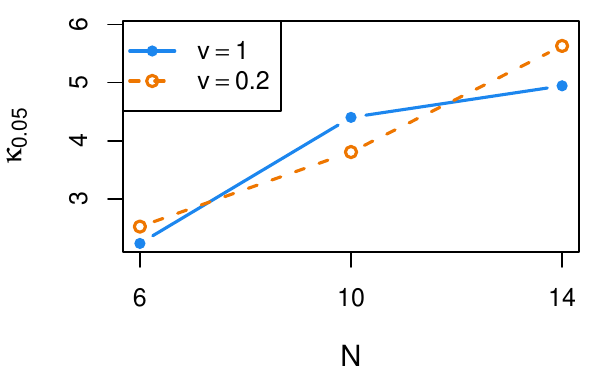}\\
    \includegraphics[width=\linewidth]{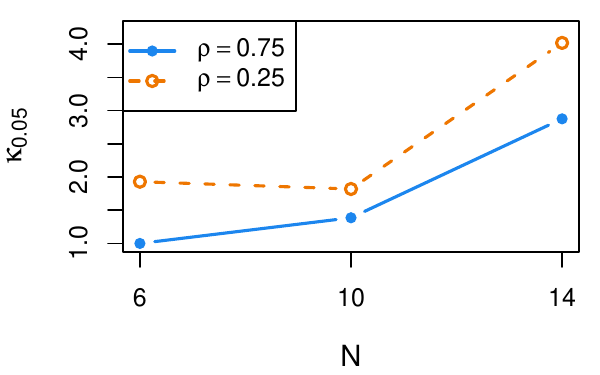}
\end{subfigure}

\caption{Monte Carlo estimates of SCS size \( \kappa_{\alpha} \) at the 95\% confidence level. Columns correspond to Sharpe ratio loss (\( -\mu/\sigma \)), mean-variance loss (\( -0.5\,\mu+\sigma^2 \)), and expected shortfall loss (\( -\mu + \sigma\,\phi(z_{0.1})/0.1 \)). Top and bottom rows correspond, respectively, to Model 1 ($v\in\{0.2, 1\}$) and Model 2 (\( \rho \in \{0.25, 0.75 \}\)).}
\label{fig:k-vs-n-panels}
\end{figure}

We examine how the number of assets \( N \) influences selection uncertainty. Figure~\ref{fig:k-vs-n-panels} shows that the expected size of the SCS increases with \( N \). As \( N \) grows, the number of feasible EWPs, \( 2^N - 1 \), rises exponentially, leading to a sharp increase in \( \kappa_\alpha \). For example, in Model 2 with \( \rho = 0.25 \) and under the Sharpe ratio loss, $\kappa_\alpha$ increases from about 8 to 231 as $N$ rises from 6 to 14, considering the $95\%$ confidence level. This growth honestly reflects the intrinsic uncertainty of a larger asset universe: as the number of feasible portfolios explodes, many display nearly identical empirical performance, making it harder to confidently identify a single superior strategy.

\subsection{Construction via Ledoit and Wolf Sharpe ratio test}

We replicate the simulation setup from Section~\ref{sec:simulation_setup}, using the Ledoit and Wolf (LW) bootstrap test for Sharpe ratio differences \citep{LW.08} to construct the SCS. The LW procedure aims to improve finite-sample accuracy of the studentized statistic $Z(\bfs, \bfs')$ by bootstrapping. The analysis is conducted using the R function \texttt{sharpeTesting} in the package \texttt{PeerPerformance} \citep{ardia2018peer}. In Table \ref{table:L&W}, we report Monte Carlo estimates of \( \kappa_\alpha \), \( \underline{\kappa}_\alpha \), and \( p_\alpha \) under Model 1 for $v \in \{0.2, 1\}$. The LW-based SCS is typically larger than that obtained via the Wald test under asymptotic normality, with differences becoming more pronounced in larger samples. For example, with \( 1-\alpha = 0.95 \) and \( v = 1 \), the LW procedure yields an expected SCS size of about 447 for \( T = 100 \) and 43 for \( T = 1000 \), compared to 295 and 25, respectively, under the asymptotic approach. This reflects a greater conservativeness of LW and potentially lower power. However, for smaller samples, the LW test achieves coverage closer to the nominal level, better controlling type I error when asymptotic approximations break down. 

\begin{table}
\centering
\setlength{\tabcolsep}{4pt}
\renewcommand{\arraystretch}{1.2}
\begin{tabular}{c c ccc ccc ccc}
\toprule
\multicolumn{1}{c}{$v$} & \multicolumn{1}{c}{$T$}
  & \multicolumn{3}{c}{$1-\alpha=0.90$}
  & \multicolumn{3}{c}{$1-\alpha=0.95$}
  & \multicolumn{3}{c}{$1-\alpha=0.99$} \\ 
\cmidrule(lr){3-5}\cmidrule(lr){6-8}\cmidrule(l){9-11}
 &  & $\kappa_\alpha$ & $p_\alpha\%$ & $\underline{\kappa}_\alpha$
   & $\kappa_\alpha$ & $p_\alpha\%$ & $\underline{\kappa}_\alpha$
   & $\kappa_\alpha$ & $p_\alpha\%$ & $\underline{\kappa}_\alpha$ \\
\midrule
\multirow{3}{*}{1}
 & 100  & \makecell{310.7\\(13.4)} & \makecell{91.0\\(1.7)} & \makecell{5.2\\(0.1)}
         & \makecell{447.4\\(15.2)} & \makecell{94.7\\(1.3)} & \makecell{6.3\\(0.1)}
         & \makecell{747.4\\(14.0)} & \makecell{100.0\\(0.0)} & \makecell{8.2\\(0.1)} \\
 & 250  & \makecell{133.8\\(6.5)} & \makecell{93.0\\(1.5)} & \makecell{3.9\\(0.1)}
         & \makecell{248.0\\(10.3)} & \makecell{97.7\\(0.9)} & \makecell{5.2\\(0.1)}
         & \makecell{515.0\\(12.9)} & \makecell{100.0\\(0.0)} & \makecell{7.1\\(0.1)} \\
 & 1000 & \makecell{25.7\\(1.5)}  & \makecell{95.7\\(1.2)} & \makecell{2.2\\(0.1)}
         & \makecell{42.5\\(2.0)}  & \makecell{98.7\\(0.7)} & \makecell{2.8\\(0.1)}
         & \makecell{118.6\\(5.0)} & \makecell{99.7\\(0.3)} & \makecell{4.3\\(0.1)} \\
\cmidrule{1-11}
\multirow{3}{*}{0.2}
 & 100  & \makecell{294.5\\(13.1)} & \makecell{84.0\\(2.1)} & \makecell{5.3\\(0.1)}
         & \makecell{441.3\\(14.5)} & \makecell{91.0\\(1.7)} & \makecell{6.4\\(0.1)}
         & \makecell{755.7\\(13.3)} & \makecell{100.0\\(0.0)} & \makecell{8.4\\(0.1)} \\
 & 250  & \makecell{164.3\\(8.6)} & \makecell{92.0\\(1.6)} & \makecell{4.4\\(0.1)}
         & \makecell{279.1\\(11.3)} & \makecell{95.3\\(1.2)} & \makecell{5.5\\(0.1)}
         & \makecell{543.3\\(13.8)} & \makecell{99.7\\(0.3)} & \makecell{7.3\\(0.1)} \\
 & 1000 & \makecell{31.9\\(1.7)}  & \makecell{96.3\\(1.1)} & \makecell{2.5\\(0.1)}
         & \makecell{51.8\\(2.6)}  & \makecell{98.7\\(0.7)} & \makecell{3.0\\(0.1)}
         & \makecell{126.1\\(5.1)} & \makecell{99.7\\(0.3)} & \makecell{4.5\\(0.1)} \\
\bottomrule
\end{tabular}
\caption{Monte Carlo estimates of SCS size \( \kappa_\alpha \), coverage probability \( p_{\alpha} \) (in \%), and lower boundary size \( \underline{\kappa}_\alpha \) for SCSs constructed using the Ledoit and Wolf bootstrap test for Sharpe ratio differences \citep{LW.08}. Monte Carlo standard errors are in parentheses.}
\label{table:L&W}
\end{table}

\section{Analysis of Cryptocurrency and Industry Portfolios}
\label{sec:examples}

In this section, we illustrate the SCS methodology by analyzing two datasets: (i) the \textit{L1-Crypto} dataset, containing daily log-returns for 16 Layer-1 cryptocurrencies from January 1, 2022, to January 1, 2025 ($T=1095$) (source: Yahoo Finance, \url{https://finance.yahoo.com}); and (ii) the French \textit{17 Industry} dataset, comprising daily returns for 17 equal-weighted U.S. industry portfolios from November 1, 2021, to October 31, 2024 ($T=755$). Equal-weighted series are used to avoid large-cap bias, providing a more balanced view of market performance (source: K. French data library \url{https://mba.tuck.dartmouth.edu/pages/faculty/ken.french/}). Tables S3 and S4 in the Supplementary Materials report additional information on the asset composition. The two datasets represent distinct market environments: the L1-Crypto data reflect high volatility and speculative dynamics, while the 17 Industry portfolios capture the stability of mature equity markets. 

We study selection uncertainty under two loss functions for each dataset: the Sharpe ratio loss, suited to the reward-to-variability focus of crypto markets, and the mean–variance loss, which is more aligned with the risk-sensitive nature of traditional equities. Point estimates of the optimal EWP obtained via exhaustive search on the L1-Crypto data are \{TRON , Mantra\} under the Sharpe ratio loss, and \{Bitcoin, TRON\} under the mean–variance loss. For the 17 Industry data, the corresponding estimates are \{Steel, Utilities\} under both the Sharpe ratio and the mean–variance losses. Next, we apply our methodology to assess the selection uncertainty surrounding these point estimates.

Table~\ref{tab:SCS_comparison} reports the global post-selection metrics defined in Section~\ref{sec:metrics} for each dataset: SCS size ($|\SCS|$), lower boundary size ($|\LB|$), Relative Multiplicity Index (RMI), loss bounds ($\hat L_0$, $\hat L_{\text{max}}$), and performance spread ($\hat \delta_\alpha$). The size of the SCS directly reflects the degree of selection uncertainty in each dataset. For example,  at the 95\% confidence level, the SCS under the Sharpe ratio for the L1-Cryptocurrency data includes 122 equally weighted portfolios (EWPs) that are statistically indistinguishable from an optimal EWP, compared to 430 in the 17 Industry dataset. This suggests that uncertainty is more contained in the L1-Cryptocurrency case. Although this may seem counterintuitive given the high volatility of cryptocurrencies, it reflects the fact that a few dominant assets drive Sharpe ratio performance, limiting the number of near-optimal configurations. In contrast, the larger number of equivalent strategies in the 17 Industry dataset indicates greater redundancy among sectors, with many combinations yielding comparable risk-adjusted performance and thus inflating the selection ambiguity. When accounting for the total size of the selection space $\Scal$ in each dataset, the RMI index confirms that selection uncertainty is lower for the L1-Cryptocurrency data (RMI$=56.68$) compared to that of the  17 Industry data (RMI$=48.54$).

As expected, higher confidence levels yield larger SCSs. The increase is substantial -- for instance, under the Sharpe ratio loss, changing the confidence level from 95\% to 99\% increases the SCS size from 122 to 8,128 for the L1-Crypto dataset, and from 430 to 10,178 for the 17 Industry dataset. The RMI decreases as the confidence level increases, reflecting the dilution of explanatory power across a larger set of selected portfolios. This highlights a fundamental trade-off: higher confidence improves coverage but reduces selectivity, making it harder to identify clearly dominant investment strategies. On the other hand, the LB set remains small -- just 12 for cryptocurrencies and 9 for equities at the 99\% confidence level under the Sharpe ratio loss. These lean allocations cannot be statistically rejected and show remarkable stability across confidence levels, suggesting a robust core of individually strong-performing assets.

\begin{table}[ht]
\centering
\begin{tabular}{@{}lcccccccc@{}}
\toprule
Dataset & $L(\mu, \sigma)$ & \((1-\alpha)\%\) &
\(|\SCS|\) &
\(|\LB|\) &
RMI (\%) &
\(\hat L_0\) (\%) &
\(\hat L_{\text{max}}\) (\%) &
\(\hat \delta_{\alpha}\) (\%) \\
\midrule
\multirow{6}{*}{L1-Crypto}
& \multirow{3}{*}{$-\mu/\sigma$}
& 90 & 20  & 3  & 72.99 & -5.09 & 1.68 & 6.77 \\
&      & 95 & 122   & 9  & 56.68 & -5.09 & 3.37 & 8.46 \\
&      & 99 & 8,128  & 12 & 18.82 & -5.09 & 3.37 & 8.46 \\
\cmidrule{2-9}
& \multirow{3}{*}{$-\mu + 0.5 \sigma^2$}
&  90 & 85   & 4  & 59.94 & 0.03 & 0.17 & 0.14 \\
& &  95 & 288  & 5  & 48.94 & 0.03 & 0.24 & 0.21 \\
& &  99 & 3,068 & 9  & 27.61 & 0.03 & 0.24 & 0.21 \\
\midrule
\multirow{6}{*}{17 Industry}
& \multirow{3}{*}{$-\mu/\sigma$}
&       90 & 115 & 5 & 59.73 & -2.96 & 2.27 & 5.23 \\
&      & 95 & 430 & 7 & 48.54 & -2.96 & 2.43 & 5.39 \\
&      & 99 & 10,178 & 9 & 21.68 & -2.96 & 5.02 & 7.98 \\
\cmidrule{2-9}
& \multirow{3}{*}{$-\mu + 0.5 \sigma^2$}
&      90 & 112     & 4  & 59.96 & -0.02 & 0.03 & 0.05 \\
&      & 95 & 408   & 7  & 48.99 & -0.02 & 0.07 & 0.09 \\
&      & 99 & 8,291 & 8 & 23.43  & -0.02 & 0.07 & 0.09 \\
\bottomrule
\end{tabular}
\caption{Post-selection metrics for the L1-Crypto and 17 Industry data under the Sharpe ratio ($-\mu/\sigma$) and mean–variance ($-\mu+0.5 \sigma^2$) loss functions, evaluated at confidence levels $(1-\alpha)\%$: SCS size ($|\SCS|$), lower boundary size ($|\LB|$), percent Relative Multiplicity Index (RMI), loss bounds ($\hat L_0$, $\hat L_{\text{max}}$), and performance spread ($\hat \delta_\alpha = \hat L_{\text{max}} - \hat L_0$).}
\label{tab:SCS_comparison}
\end{table}

\begin{figure}[ht]
\centering

\begin{subfigure}[t]{0.49\textwidth}
  \centering
  \includegraphics[width=\textwidth]{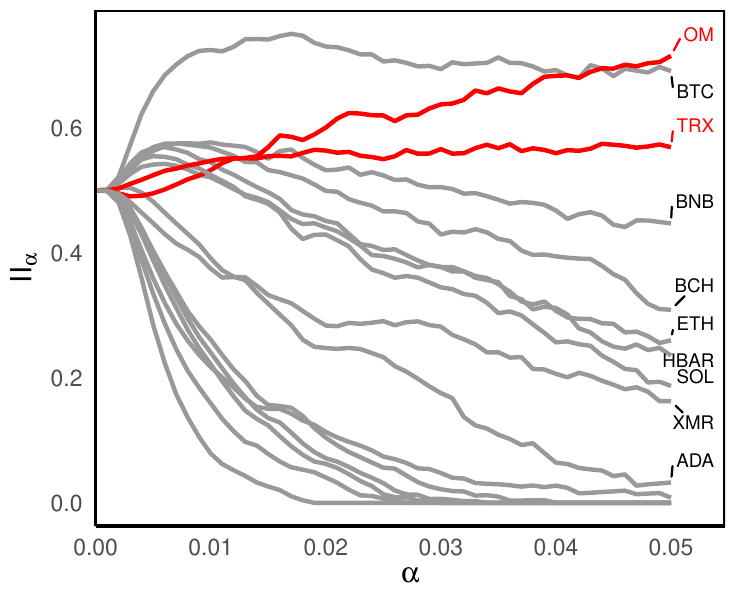}
\end{subfigure}
\begin{subfigure}[t]{0.49\textwidth}
  \centering
  \includegraphics[width=\textwidth]{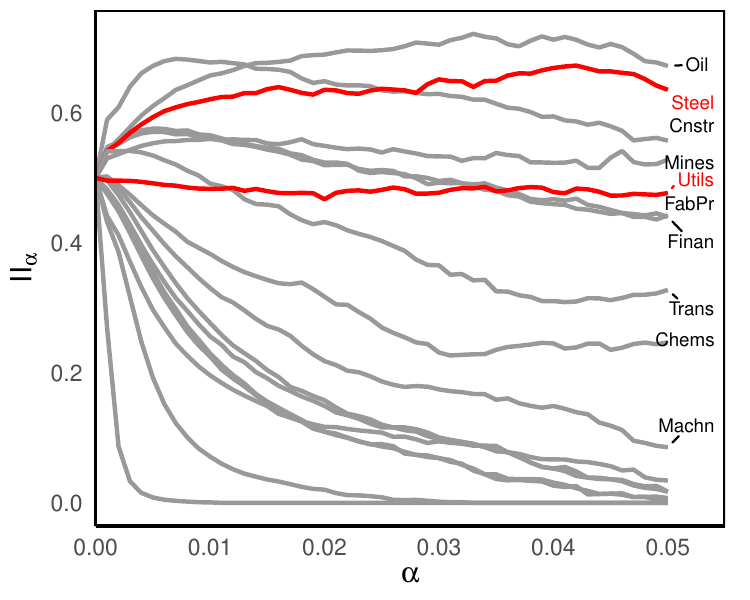}
\end{subfigure}
\caption{Inclusion importance ($\text{II}_\alpha$) as a function of $\alpha$ for the L1-Crypto dataset under the Sharpe ratio loss (left) and for the 17 Industry dataset under the mean-variance loss (right). 
Red lines mark assets included in the empirical optimal portfolio. 
Labels are reported for the 10 assets with the largest $\text{II}_{\alpha}$ value at $\alpha=0.05$.}
\label{fig:II_profiles}
\end{figure}

In Figure~\ref{fig:II_profiles}, we plot the marginal inclusion importance (\(\text{II}_\alpha\)) for each asset over \(\alpha \in [0,0.05]\) for the L1-Crypto and 17-Industry datasets, under the Sharpe ratio and mean–variance loss, respectively. Some assets maintain high and stable inclusion importance as \(\alpha\) increases, indicating frequent selection in top-performing portfolios across confidence levels. Others show a rapid decline, with their trajectories quickly clustering below the rest, signaling weak and unstable contributions to portfolio performance. In the L1-Crypto data, Bitcoin (BTC) and Mantra (OM) consistently exhibit large inclusion importance, reflecting their central role under the Sharpe ratio criterion. In the 17-Industry data, Oil \& Gas, Steel, and Construction Materials stand out dominate the other sectors in terms of inclusion importance under the mean–variance criterion.

These plots illustrate a key strength of our approach: important assets may be excluded from the empirical optimum due to sampling variability, yet still emerge as influential within the full Selection Confidence Set (SCS). In the L1-Crypto data under the mean-variance loss, the two assets composing the empirical optimum (TRX and OM) consistently receive high inclusion importance across all confidence levels. However, other key assets -- such as Bitcoin (BTC) -- are not part of the best empirical EWP despite their relevance. Similar considerations apply to the 17 Industry data. While Steel and Utilities are part of the empirically optimal portfolio under the minimum-loss, there are  several sectors with high inclusion importance --  such as Oil \& Gas, Construction Materials Drugs and Mines -- entirely ignored by the single selected EWP. By examining the entire SCS, we identify systematically important assets that a single optimal EWP would overlook, highlighting how the full SCS reveals structural relevance of assets beyond a single empirically selected EWP.

\begin{figure}[ht]
\centering
\begin{subfigure}[t]{0.48\textwidth}
  \centering
  \includegraphics[width=\textwidth]{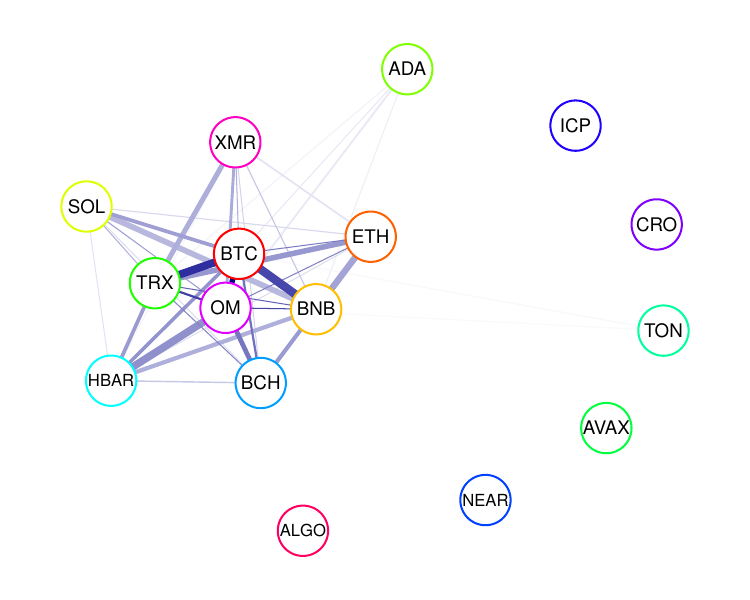}
\end{subfigure}
\hfill
\begin{subfigure}[t]{0.48\textwidth}
  \centering
  \includegraphics[width=\textwidth]{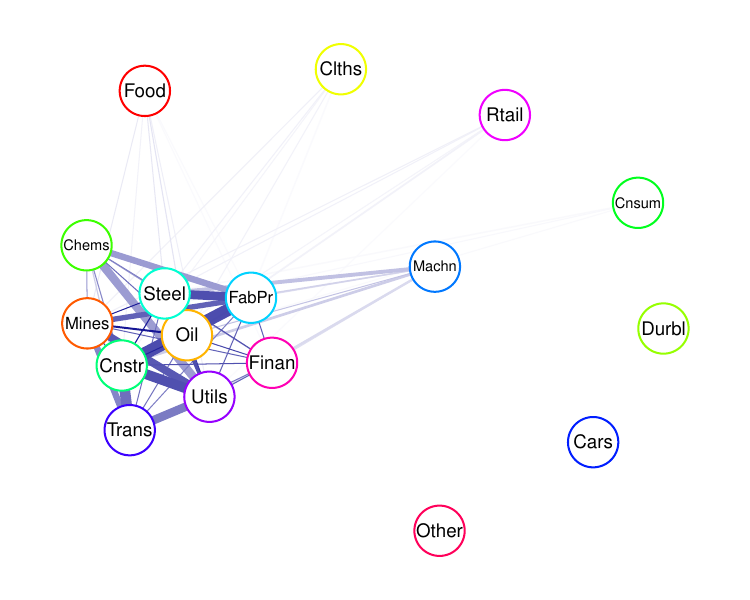}
\end{subfigure}
\caption{Co-inclusion importance ($\text{CII}_\alpha$) graphs at the 95\% confidence level for the L1-Crypto dataset under the Sharpe ratio loss (left) and for the 17 Industry dataset under the mean-variance loss (right). The edges represent pairs of assets such that $\text{CII}_\alpha>0.01$; thicker edges correspond to larger co-inclusion values.}\label{fig:CII_networks}
\end{figure}

Figure~\ref{fig:CII_networks} shows graphs representing the co-inclusion importance metric (\(\text{CII}_\alpha\)) at the 95\% confidence level. Both datasets exhibit dense connectivity, indicating that many asset pairs frequently co-appear in EWPs within the SCS. For the L1-Crypto data, a few assets such as Bitcoin (BTC) and Mantra (OM) are highly central with many edges concentrated around these assets, reflecting their frequent joint inclusion in high-Sharpe portfolios. For the 17-Industry data, Oil \& Gas, Steel, Utilities, and Construction Materials dominate in centrality, indicating that these sectors are consistently selected alongside others under the mean–variance criterion. This pattern reflects strong complementarities, where their inclusion typically enhances overall portfolio performance. By contrast, isolated nodes represent assets with limited complementarity, contributing little when combined with others and often playing a marginal or standalone role in portfolio construction.

Overall, although cryptocurrencies and industries differ markedly in their return and volatility profiles, both yield similarly large confidence sets -- highlighting the inherent uncertainty in identifying a single optimal EWP. Rather than relying solely on a unique selection, our method complements it with richer information by uncovering a range of statistically indistinguishable alternatives and systematically important assets, offering a more robust foundation for portfolio decisions.

\section{Conclusion} 
\label{sec:Conclusion}

Statistically grounded confidence sets provide a rigorous framework for addressing a longstanding but often unquantified issue in portfolio theory -- the abundance of nearly optimal solutions arising from uncertainty in expected return and risk estimates. While estimation risk for a single portfolio is well documented, our SCS approach addresses statistical uncertainty by identifying all portfolios that are indistinguishable from the true optimum at a specified confidence level. Using a flexible framework compatible with a broad class of performance losses, we establish asymptotic coverage  and show that the expected size of the SCS depends on the signal-to-noise ratios of competing portfolios—shrinking exponentially in $T$ as identification strengthens.

We remark that the test used to construct the SCS is asymptotically most powerful among those based on normally distributed estimators (see \citet{lehmann2005testing}). As such, the resulting SCS serves as a best-case benchmark for our pairwise approach: its size reflects the minimal inclusion needed to control type I error at level $\alpha$. Despite this optimality, the SCS can still be large when uncertainty is high, particularly when performance differences are faint. This largeness is not a flaw, but a statistical necessity, reflecting the indistinguishability between the true optimum and many near-optimal portfolios. Without additional structural information on the space of feasible portfolios $\Scal$, such uncertainty cannot be further reduced. Future work may explore incorporating shape constraints, economic priors, or regularization to refine the SCS without compromising coverage.

While our analysis assumes a fixed number of assets $N$, extending the methodology to high-dimensional settings presents nontrivial theoretical and computational challenges. Nevertheless, the asymptotic upper bound in Equation~\eqref{eq:scs_size_bound} shows that the SCS size decays exponentially with $T$. This suggests that, under appropriate sparsity or separability conditions on the loss function, selection complexity may remain tractable even as $N$ increases, offering a promising direction for both theoretical and computational extensions to high-dimensional portfolio problems.

Our numerical experiments confirm the theoretical behavior of the SCS, with coverage approaching the nominal level as sample size increases. However, in high-dimensional settings with weakly identified loss differentials, coverage may fall short of $1-\alpha$ due to limitations of asymptotic approximations.  Based on the observed finite sample behavior, future work may consider bootstrap or higher-order corrections for \( Z(\bfs,\hat{\bfs}_0) \), or ad hoc tests tailored to specific loss functions. However, we observed that while the bootstrapped test of \citep{LW.08} slightly improves coverage in small samples, it does not fundamentally reduce selection uncertainty under weak signals and often yields larger SCSs than those obtained using asymptotic normality (see Table~\ref{table:L&W}). The proof of Proposition~\ref{prop:scs_coverage} shows that SCS validity relies on the consistency of the selection procedure—specifically, a high probability of correct selection \( \mathbb{P}(\hat{\bfs}_0 \in \mathcal{S}_0) \). While this holds asymptotically, finite-sample performance may suffer, particularly in high-dimensional or weak-signal settings, leading to reduced coverage. Incorporating uncertainty in \( \hat{\bfs}_0 \) -- e.g., by adjusting the variance estimator or using bootstrap methods -- could improve the finite-sample accuracy of the SCS.

Through our real-data illustrations, we demonstrate that selection uncertainty is not an abstract concept but a concrete feature of portfolio construction. In both cryptocurrency and equity markets, a large number of EWPs cannot be statistically distinguished from the optimum under standard loss functions. This finding highlights the limitations of approaches relying solely on point estimates and motivates a shift toward distributional characterizations of portfolio selection. The diagnostic tools introduced (such as the RMI, loss intervals, inclusion importance, and co-inclusion networks) are interpretable metrics to assess the stability and structure of the SCS. These tools provide practitioners with a principled basis for evaluating portfolio robustness, guiding diversification choices, and understanding the extent of selection ambiguity. 

\section*{Disclosure statement}  
The authors declare that there are no conflicts of interest to disclose.  

\section*{Data Availability Statement}  
The data that support the findings of this study are openly available in Harvard Dataverse at \url{https://doi.org/10.7910/DVN/VG0FU6}, reference number V1.

\bibliographystyle{abbrvnat}
\bibliography{biblio}

\begin{thebibliography}{38}
\providecommand{\natexlab}[1]{#1}
\providecommand{\url}[1]{\texttt{#1}}
\expandafter\ifx\csname urlstyle\endcsname\relax
  \providecommand{\doi}[1]{doi: #1}\else
  \providecommand{\doi}{doi: \begingroup \urlstyle{rm}\Url}\fi

\bibitem[Amendola et~al.(2020)Amendola, Braione, Candila, and Storti]{AM.20}
A.~Amendola, M.~Braione, V.~Candila, and G.~Storti.
\newblock A model confidence set approach to the combination of multivariate
  volatility forecasts.
\newblock \emph{International Journal of Forecasting}, 36\penalty0
  (3):\penalty0 873--891, 2020.
\newblock \doi{10.1016/j.ijforecast.2019.10.001}.

\bibitem[Andrews(1991)]{andrews1991heteroskedasticity}
D.~W. Andrews.
\newblock Heteroskedasticity and autocorrelation consistent covariance matrix
  estimation.
\newblock \emph{Econometrica: Journal of the Econometric Society}, pages
  817--858, 1991.

\bibitem[Ardia and Boudt(2018)]{ardia2018peer}
D.~Ardia and K.~Boudt.
\newblock The peer performance ratios of hedge funds.
\newblock \emph{Journal of Banking \& Finance}, 87:\penalty0 351--368, 2018.

\bibitem[Asawa(2021)]{A.21}
Y.~S. Asawa.
\newblock Modern machine learning solutions for portfolio selection.
\newblock \emph{IEEE Engineering Management Review}, 50\penalty0 (1):\penalty0
  94--112, 2021.

\bibitem[Bauder et~al.(2021)Bauder, Bodnar, Parolya, and Schmid]{B.21}
D.~Bauder, T.~Bodnar, N.~Parolya, and W.~Schmid.
\newblock Bayesian mean–variance analysis: Optimal portfolio selection under
  parameter uncertainty.
\newblock \emph{Quantitative Finance}, 21\penalty0 (2):\penalty0 221--242,
  2021.

\bibitem[Bodnar et~al.(2025)Bodnar, Bodnar, and Niklasson]{Bod.25}
O.~Bodnar, T.~Bodnar, and V.~Niklasson.
\newblock Incorporating different sources of information for bayesian optimal
  portfolio selection.
\newblock \emph{Journal of Business \& Economic Statistics}, 43\penalty0
  (2):\penalty0 365--377, 2025.
\newblock \doi{10.1080/07350015.2024.2379361}.

\bibitem[Bodnar et~al.(2022{\natexlab{a}})Bodnar, Lindholm, Niklasson, and
  Thorsén]{BO.22}
T.~Bodnar, M.~Lindholm, V.~Niklasson, and E.~Thorsén.
\newblock Bayesian portfolio selection using var and cvar.
\newblock \emph{Applied Mathematics and Computation}, 427:\penalty0 127120,
  2022{\natexlab{a}}.

\bibitem[Bodnar et~al.(2022{\natexlab{b}})Bodnar, Okhrin, and Parolya]{Bod.22}
T.~Bodnar, Y.~Okhrin, and N.~Parolya.
\newblock Optimal shrinkage-based portfolio selection in high dimensions.
\newblock \emph{Journal of Business \& Economic Statistics}, 41\penalty0
  (1):\penalty0 140--156, 2022{\natexlab{b}}.
\newblock \doi{10.1080/07350015.2021.2004897}.

\bibitem[Bodnar et~al.(2024)Bodnar, Parolya, and Thorsén]{BO.24}
T.~Bodnar, N.~Parolya, and E.~Thorsén.
\newblock Two is better than one: Regularized shrinkage of large minimum
  variance portfolios.
\newblock \emph{Journal of Machine Learning Research}, 25\penalty0
  (173):\penalty0 1--32, 2024.

\bibitem[Caparrini et~al.(2024)Caparrini, Arroyo, and Escayola~Mansilla]{CA.24}
A.~Caparrini, J.~Arroyo, and J.~Escayola~Mansilla.
\newblock S\&p 500 stock selection using machine learning classifiers: A look
  into the changing role of factors.
\newblock \emph{Research in International Business and Finance}, 70:\penalty0
  102336, June 2024.
\newblock \doi{10.1016/j.ribaf.2024.102336}.

\bibitem[Chen et~al.(2021)Chen, Zhang, Mehlawat, and Jia]{C.21}
W.~Chen, H.~Zhang, M.~K. Mehlawat, and L.~Jia.
\newblock Mean–variance portfolio optimization using machine learning-based
  stock price prediction.
\newblock \emph{Applied Soft Computing}, 100:\penalty0 106943, 2021.

\bibitem[Chen et~al.(2023)Chen, Zhang, and Weng]{CH.23}
Z.~Chen, L.~Zhang, and C.~Weng.
\newblock Does climate policy uncertainty affect chinese stock market
  volatility?
\newblock \emph{International Review of Economics \& Finance}, 84:\penalty0
  369--381, 2023.
\newblock \doi{10.1016/j.iref.2022.11.030}.

\bibitem[Chopra and Ziemba(2013)]{C.13}
V.~K. Chopra and W.~T. Ziemba.
\newblock \emph{The Effect of Errors in Means, Variances, and Covariances on
  Optimal Portfolio Choice}, pages 365--373.
\newblock 2013.

\bibitem[DeMiguel et~al.(2009)DeMiguel, Garlappi, and Uppal]{DM.09}
V.~DeMiguel, L.~Garlappi, and R.~Uppal.
\newblock Optimal versus naive diversification: How inefficient is the 1/n
  portfolio strategy?
\newblock \emph{The review of Financial studies}, 22\penalty0 (5):\penalty0
  1915--1953, 2009.

\bibitem[DeMiguel et~al.(2013)DeMiguel, Martin-Utrera, and Nogales]{DM.13}
V.~DeMiguel, A.~Martin-Utrera, and F.~J. Nogales.
\newblock Size matters: Optimal calibration of shrinkage estimators for
  portfolio selection.
\newblock \emph{Journal of Banking \& Finance}, 37\penalty0 (8):\penalty0
  3018--3034, 2013.

\bibitem[Ding et~al.(2021)Ding, Li, and Zheng]{Din.21}
Y.~Ding, Y.~Li, and X.~Zheng.
\newblock High dimensional minimum variance portfolio estimation under
  statistical factor models.
\newblock \emph{Journal of Econometrics}, 222\penalty0 (1):\penalty0 502--515,
  2021.
\newblock \doi{10.1016/j.jeconom.2020.07.013}.

\bibitem[Efron et~al.(2004)Efron, Hastie, Johnstone, and Tibshirani]{E.04}
B.~Efron, T.~Hastie, I.~Johnstone, and R.~Tibshirani.
\newblock Least angle regression.
\newblock \emph{The Annals of Statistics}, 32:\penalty0 407–499, 2004.

\bibitem[Fama and French(1992)]{FF.92}
E.~Fama and K.~French.
\newblock The cross-section of expected stock returns.
\newblock \emph{The Journal of Finance}, 47\penalty0 (2):\penalty0 427--465,
  1992.

\bibitem[Ferrari and Yang(2015)]{FY.15}
D.~Ferrari and Y.~Yang.
\newblock Condence sets for model selection by f-testing.
\newblock \emph{Statistica Sinica}, pages 1637--1658, 2015.

\bibitem[Goto and Xu(2015)]{GO.15}
S.~Goto and Y.~Xu.
\newblock Improving mean variance optimization through sparse hedging
  restrictions.
\newblock \emph{Journal of Financial and Quantitative Analysis}, 50\penalty0
  (6):\penalty0 1415--1441, 2015.
\newblock \doi{10.1017/S0022109015000526}.

\bibitem[Hansen et~al.(2011)Hansen, Lunde, and J.M.]{H.11}
P.~Hansen, A.~Lunde, and N.~J.M.
\newblock The model confidence set.
\newblock \emph{Econometrica}, 79\penalty0 (2):\penalty0 453--497, 2011.

\bibitem[Jiang et~al.(2021)Jiang, Fei, Liu, Roeder, Lafferty, Wasserman, Li,
  and Zhao]{huge}
H.~Jiang, X.~Fei, H.~Liu, K.~Roeder, J.~Lafferty, L.~Wasserman, X.~Li, and
  T.~Zhao.
\newblock \emph{huge: High-Dimensional Undirected Graph Estimation}, 2021.
\newblock URL \url{https://CRAN.R-project.org/package=huge}.
\newblock R package version 1.3.5.

\bibitem[Kritzman et~al.(2010)Kritzman, Page, and Turkington]{K.10}
M.~Kritzman, S.~Page, and D.~Turkington.
\newblock In defense of optimization: the fallacy of 1/n.
\newblock \emph{Financial Analysts Journal}, 66:\penalty0 31--39, 2010.

\bibitem[Ledoit and Wolf(2008)]{LW.08}
O.~Ledoit and M.~Wolf.
\newblock Robust performance hypothesis testing with the sharpe ratio.
\newblock \emph{Journal of Empirical Finance}, 15\penalty0 (5):\penalty0
  850--859, 2008.

\bibitem[Ledoit and Wolf(2011)]{LW.11}
O.~Ledoit and M.~Wolf.
\newblock Robust performances hypothesis testing with the variance.
\newblock \emph{Wilmott}, \penalty0 (55):\penalty0 86--89, 2011.

\bibitem[Ledoit and Wolf(2018)]{LW.18}
O.~Ledoit and M.~Wolf.
\newblock Robust performance hypothesis testing with smooth functions of
  population moments.
\newblock University of Zurich, Department of Economics, Working Paper, (305),
  2018.

\bibitem[Lee(2020)]{L.20}
S.~I. Lee.
\newblock Deeply equal-weighted subset portfolios.
\newblock \emph{arXiv preprint arXiv:2006.14402}, 2020.

\bibitem[Lehmann and Romano(2005)]{lehmann2005testing}
E.~L. Lehmann and J.~P. Romano.
\newblock \emph{Testing statistical hypotheses}.
\newblock Springer, 2005.

\bibitem[Leippold and Yang(2023)]{Lei.23}
M.~Leippold and H.~Yang.
\newblock Mixed‐frequency predictive regressions with parameter learning.
\newblock \emph{Journal of Forecasting}, 42\penalty0 (8):\penalty0 1955--1972,
  2023.
\newblock \doi{10.1002/for.2999}.

\bibitem[Liang et~al.(2022)Liang, Umar, Ma, and Huynh]{LI.22}
C.~Liang, M.~Umar, F.~Ma, and T.~L.~D. Huynh.
\newblock Climate policy uncertainty and world renewable energy index
  volatility forecasting.
\newblock \emph{Technological Forecasting and Social Change}, 182:\penalty0
  121810, 2022.
\newblock \doi{10.1016/j.techfore.2022.121810}.

\bibitem[Malladi and Fabozzi(2017)]{M.17}
R.~Malladi and F.~J. Fabozzi.
\newblock Equal-weighted strategy: Why it outperforms value-weighted
  strategies? theory and evidence.
\newblock \emph{Journal of Asset Management}, 18:\penalty0 188--208, 2017.

\bibitem[Markowitz(1952)]{M.52}
H.~Markowitz.
\newblock Portfolio selection.
\newblock \emph{The Journal of Finance}, 7\penalty0 (1):\penalty0 77--91, 1952.

\bibitem[Memmel(2003)]{Me.03}
C.~Memmel.
\newblock Performance hypothesis testing with the sharpe ratio.
\newblock \emph{Finance Letters}, 1\penalty0 (1):\penalty0 21--23, 2003.
\newblock URL \url{https://ssrn.com/abstract=412588}.

\bibitem[Paiva et~al.(2019)Paiva, Cardoso, Hanaoka, and Duarte]{P.19}
F.~D. Paiva, R.~T.~N. Cardoso, G.~P. Hanaoka, and W.~M. Duarte.
\newblock Decision-making for financial trading: A fusion approach of machine
  learning and portfolio selection.
\newblock \emph{Expert Systems with Applications}, 115:\penalty0 635--655,
  2019.

\bibitem[Plyakha et~al.(2015)Plyakha, Uppal, and Vilkov]{P.15}
Y.~Plyakha, R.~Uppal, and G.~Vilkov.
\newblock Why do equal-weighted portfolios outperform value-weighted
  portfolios.
\newblock \emph{SSRN Electronic Journal}, 2015.

\bibitem[Ross(1976)]{R.76}
S.~A. Ross.
\newblock The arbitrage theory of capital asset pricing.
\newblock \emph{Journal of Economic Theory}, 13\penalty0 (3):\penalty0
  341--360, 1976.

\bibitem[Sharpe(1964)]{S.64}
W.~F. Sharpe.
\newblock Capital asset prices: A theory of market equilibrium under conditions
  of risk.
\newblock \emph{The Journal of Finance}, 19\penalty0 (3):\penalty0 425--442,
  1964.

\bibitem[Zheng et~al.(2019)Zheng, Ferrari, and Yang]{Z.19}
C.~Zheng, D.~Ferrari, and Y.~Yang.
\newblock Model selection confidence sets by likelihood ratio testing.
\newblock \emph{Statistica Sinica}, 29\penalty0 (2):\penalty0 827--851, 2019.

\end{thebibliography}

\section{Appendix: Proofs of Propositions}

\begin{proof}[Proof of Proposition \ref{prop:scs_coverage}]
Fix any \( \bfs_0 \in \mathcal{S}_0 \).  Using the law of total probability we obtain the following lower bound for inclusion probability \( \mathbb{P}(\bfs_0 \in \SCS) \)
\begin{equation}\label{eq:ltp_bound}
\mathbb{P}(\bfs_0 \in \SCS) 
= \mathbb{P}(Z(\bfs_0, \hat{\bfs}_0) \le q_{1-\alpha})  \geq \mathbb{P}(Z(\bfs_0, \hat{\bfs}_0) \le  q_{1-\alpha} \mid \hat{\bfs}_0 \in \mathcal{S}_0)\, \mathbb{P}(\hat{\bfs}_0 \in \mathcal{S}_0). 
\end{equation}

{\it Step 1.} We first show $\mathbb{P}(\hat{\bfs}_0 \in \mathcal{S}_0) \to 1$ as $T \to \infty$. Let \( L_Z(\mathbf{s}) := L(\hat\mu_{\mathbf{s}}, \hat\sigma^2_{\mathbf{s}}) \) and \( L(\mathbf{s}) := L(\mu_{\mathbf{s}}, \sigma^2_{\mathbf{s}}) \). By Assumption (A2), for each fixed \( \mathbf{s} \in \mathcal{S} \) 
$
(\hat\mu_{\mathbf{s}}, \hat\sigma^2_{\mathbf{s}}) \overset{p}{\to} (\mu_{\mathbf{s}}, \sigma^2_{\mathbf{s}})$,
and by continuity of \( L \), which implies 
$L_Z(\mathbf{s}) \overset{p}{\to} L(\mathbf{s})$ for all $\mathbf{s} \in \mathcal{S}$ and
$
\max_{\mathbf{s} \in \mathcal{S}} |L_Z(\mathbf{s}) - L(\mathbf{s})| \overset{p}{\to} 0$.
Since \( \mathcal{S} \) is finite and \( L(\mathbf{s}) > L(\mathbf{s}_0) \) for all \( \mathbf{s} \notin \mathcal{S}_0 \), define the minimal separation margin between any non-optimal and optimal selections as: $
\epsilon := \min_{\mathbf{s} \notin \mathcal{S}_0} \min_{\mathbf{s}_0 \in \mathcal{S}_0} \left\{ L(\mathbf{s}) - L(\mathbf{s}_0) \right\} > 0$ ,
For sufficiently large \( T \), with probability tending to one, we have $
|L_Z(\mathbf{s}) - L(\mathbf{s})| < \epsilon/4 \quad \text{for all } \mathbf{s} \in \mathcal{S}$. Now fix any \( \mathbf{s}_0 \in \mathcal{S}_0 \) and any \( \mathbf{s} \notin \mathcal{S}_0 \). On this high-probability event, we have:
\begin{align*}
L_Z(\mathbf{s}) &> L(\mathbf{s}) - \epsilon/4 \ge L(\mathbf{s}_0) + \epsilon - \epsilon/4 = L(\mathbf{s}_0) + 3\epsilon/4  >  L_Z(\mathbf{s}_0) + \epsilon/2.
\end{align*}
This shows that, with probability tending to one, all non-minimizers have strictly larger empirical loss than any population minimizer. Therefore, Step 1 is complete.

{\it Step 2.}
We now show that under the event \( \{ \hat{\bfs}_0 \in \mathcal{S}_0 \} \), the test statistic
\[
Z(\bfs_0, \hat{\bfs}_0) = \frac{{L}_Z(\bfs_0) - {L}_Z(\hat{\bfs}_0)}{\hat{\tau}(\bfs_0, \hat{\bfs}_0)/\sqrt{T}}.
\]
 converges in distribution to a standard normal under the null hypothesis \( H_0: L(\bfs_0) = L(\hat{\bfs}_0) \).  For fixed $\hat \bfs_0$, by Assumption (A2), the empirical moment estimators are jointy asymptotically normal. Hence, by the delta method, $ \sqrt{T} \tau(\bfs_0, \hat{\bfs}_0) [{L}_Z(\bfs_0) - {L}_Z(\hat{\bfs}_0)] \overset{d}{\to} \mathcal{N}(0,1)$. By Assumption (A3), the estimated standard error in the denominator satisfies $\hat{\tau}(\bfs_0, \hat{\bfs}_0) \overset{p}{\to} \tau(\bfs_0, \hat{\bfs}_0)$. By Slutsky's theorem, the test statistic satisfies
$Z(\bfs_0, \hat{\bfs}_0) \overset{d}{\to} \mathcal{N}(0,1)$ under the null.  Therefore, the probability of not rejecting  under the null satisfies
\[
\mathbb{P}(Z(\bfs_0, \hat{\bfs}_0) \le q_{1-\alpha} \mid \hat{\bfs}_0 \in \mathcal{S}_0) \to 1 - \alpha.
\]
From Step 1 and Step 2, taking the limit for $T\to \infty$ for the right hand side in \eqref{eq:ltp_bound} gives $(1-\alpha)$, which concludes the proof.
\end{proof}

\begin{proof}[Proof of Proposition~\ref{prop:asymptotic_scs_size}]
Let \( \hat{\delta}(\bfs) = \hat{L}(\bfs) - \hat{L}(\hat{\bfs}_0) \) denote the empirical loss difference, and let \( \delta(\bfs) = L(\bfs) - L_0 \) denote the population counterpart, where \( L_0 = L(\bfs_0) \) for some \( \bfs_0 \in \Scal_0 \). Conditioning on the event \( \hat{\bfs}_0 = \bfs_0 \in \Scal_0 \),  the Delta method and  Assumptions (A1)--(A2), imply that the test statistic satisfies 
\[
Z(\bfs, \hat{\bfs}_0) \mid \hat{\bfs}_0 =\bfs_0\overset{d}{\to} \mathcal{N}\left( \frac{\delta(\bfs; {\bfs}_0)}{\tau(\bfs; {\bfs}_0)}, 1 \right), \quad \text{as } T \to \infty.
\]
Therefore, conditioning on $\hat{\bfs}_0 = \bfs_0$, the expected size of the selection confidence set satisfies :
\[
\mathbb{E}[|\SCS|] = \sum_{\bfs \in \Scal_0} \mathbb{P}(Z(\bfs, \bfs_0) \leq q_{1 - \alpha}) + \sum_{\bfs \notin \Scal_0} \mathbb{P}(Z(\bfs, \bfs_0) \leq q_{1 - \alpha}) + o(1).
\]
For \( \bfs \in \Scal_0 \), we have \( \delta(\bfs; \bfs_0) = 0 \), so $Z(\bfs, \bfs_0) \overset{d}{\to} \mathcal{N}(0,1)$ and $\mathbb{P}(Z(\bfs, \bfs_0) \leq q_{1 - \alpha}) = (1 - \alpha) + o(1)$. For \( \bfs \notin \Scal_0 \), the population loss difference is positive, and the test statistic satisfies
$\sqrt{T}Z(\bfs, \bfs_0) \overset{d}{\to} \mathcal{N}(\gamma(\bfs), 1)$ where $\gamma(\bfs) := {\delta(\bfs)}/{\tau(\bfs)}$.
Combining both parts, we obtain:
\begin{equation}\label{eq:size_null}
\mathbb{E}[|\SCS| \mid \hat{\bfs}_0 \in \Scal_0] = |\Scal_0|(1 - \alpha) + \sum_{\bfs \notin \Scal_0} \Phi(q_{1 - \alpha} - \sqrt{T} \gamma(\bfs)) + o(1).
\end{equation}
By the law of total expectation,
\begin{equation}\label{eq:lte}
\mathbb{E}[|\SCS|] = \mathbb{E}[|\SCS| \mid \hat{\bfs}_0 \in \Scal_0] \cdot \mathbb{P}(\hat{\bfs}_0 \in \Scal_0) + \mathbb{E}[|\SCS| \mid \hat{\bfs}_0 \notin \Scal_0] \cdot \mathbb{P}(\hat{\bfs}_0 \notin \Scal_0).
\end{equation}
We already showed in the proof of Proposition \ref{prop:scs_coverage} that  $\mathbb{P}(\hat{\bfs}_0 \in \Scal_0) \to 1$ under (A1). Thus, the first term in \eqref{eq:lte} is $\mathbb{E}[|\SCS| \mid \hat{\bfs}_0 \in \Scal_0] (1+ o(1))$. Since \( \SCS \subseteq \Scal \), its size is uniformly bounded above by \( |\Scal| < \infty \), and we already showed in the proof of Proposition \ref{prop:scs_coverage} that  \( \mathbb{P}(\hat{\bfs}_0 \notin \Scal_0) \to 0 \). The second term of \eqref{eq:lte} is upper bounded by $|\Scal| \cdot \mathbb{P}(\hat{\bfs}_0 \notin \Scal_0) = o(1)$. Hence, by substituting \eqref{eq:size_null} into (\eqref{eq:lte}) we conclude 
\begin{equation}
\mathbb{E}[|\SCS|] = |\Scal_0|(1 - \alpha) + \sum_{\bfs \notin \Scal_0} \Phi(q_{1 - \alpha} - \sqrt{T} \gamma(\bfs))  + o(1),
\end{equation}
and concludes the proof.
\end{proof}

\end{document}